\documentclass[
 aps,prl,showpacs,superscriptaddress,numerical,twocolumn,amsmath,amssymb,floatfix,
 reprint%
]{revtex4-1}
\usepackage{xcolor}
\usepackage[
	pdffitwindow=true,
	colorlinks=true,
	frenchlinks=false,
        linkcolor=blue,
	anchorcolor=blue,
        citecolor=blue,
        filecolor=blue,
        urlcolor=blue,
        bookmarks=true,
        bookmarksopen=true,
	bookmarksnumbered=true, 
        bookmarksopenlevel=1,
        plainpages=false,
	pdfpagelayout=OnePage,
        pdfpagelabels=true,
	breaklinks
]{hyperref}
\usepackage[per-mode=symbol,separate-uncertainty]{siunitx}
\usepackage{graphicx}
\usepackage{dcolumn}
\usepackage{bm}
\usepackage[english]{babel}
\usepackage{cases}
\usepackage{braket}

\begin{document}
\preprint{AIP/123-QED}

\title[]{Parity-engineered light-matter interaction}

\author{J.~Goetz}
\email[]{jan.goetz@wmi.badw.de}
\altaffiliation[present address:~]{QCD Labs, Department of Applied Physics, Aalto University, Aalto, Finland}
\affiliation{Walther-Mei{\ss}ner-Institut, Bayerische Akademie der Wissenschaften, 85748 Garching, Germany }
\affiliation{Physik-Department, Technische Universit\"{a}t M\"{u}nchen, 85748 Garching, Germany}
\author{F.~Deppe}
\affiliation{Walther-Mei{\ss}ner-Institut, Bayerische Akademie der Wissenschaften, 85748 Garching, Germany }
\affiliation{Physik-Department, Technische Universit\"{a}t M\"{u}nchen, 85748 Garching, Germany}
\affiliation{Nanosystems Initiative Munich (NIM), Schellingstra{\ss}e 4, 80799 M\"{u}nchen, Germany}
\author{K.~G.~Fedorov}
\affiliation{Walther-Mei{\ss}ner-Institut, Bayerische Akademie der Wissenschaften, 85748 Garching, Germany }
\affiliation{Physik-Department, Technische Universit\"{a}t M\"{u}nchen, 85748 Garching, Germany}
\author{P.~Eder}
\affiliation{Walther-Mei{\ss}ner-Institut, Bayerische Akademie der Wissenschaften, 85748 Garching, Germany }
\affiliation{Physik-Department, Technische Universit\"{a}t M\"{u}nchen, 85748 Garching, Germany}
\affiliation{Nanosystems Initiative Munich (NIM), Schellingstra{\ss}e 4, 80799 M\"{u}nchen, Germany}
\author{M.~Fischer}
\affiliation{Walther-Mei{\ss}ner-Institut, Bayerische Akademie der Wissenschaften, 85748 Garching, Germany }
\affiliation{Physik-Department, Technische Universit\"{a}t M\"{u}nchen, 85748 Garching, Germany}
\affiliation{Nanosystems Initiative Munich (NIM), Schellingstra{\ss}e 4, 80799 M\"{u}nchen, Germany}
\author{S.~Pogorzalek}
\affiliation{Walther-Mei{\ss}ner-Institut, Bayerische Akademie der Wissenschaften, 85748 Garching, Germany }
\affiliation{Physik-Department, Technische Universit\"{a}t M\"{u}nchen, 85748 Garching, Germany}
\author{E.~Xie}
\affiliation{Walther-Mei{\ss}ner-Institut, Bayerische Akademie der Wissenschaften, 85748 Garching, Germany }
\affiliation{Physik-Department, Technische Universit\"{a}t M\"{u}nchen, 85748 Garching, Germany}
\affiliation{Nanosystems Initiative Munich (NIM), Schellingstra{\ss}e 4, 80799 M\"{u}nchen, Germany}
\author{A.~Marx}
\affiliation{Walther-Mei{\ss}ner-Institut, Bayerische Akademie der Wissenschaften, 85748 Garching, Germany }
\author{R.~Gross}
\email[]{rudolf.gross@wmi.badw.de}
\affiliation{Walther-Mei{\ss}ner-Institut, Bayerische Akademie der Wissenschaften, 85748 Garching, Germany }
\affiliation{Physik-Department, Technische Universit\"{a}t M\"{u}nchen, 85748 Garching, Germany}
\affiliation{Nanosystems Initiative Munich (NIM), Schellingstra{\ss}e 4, 80799 M\"{u}nchen, Germany}

\date{\today}

\begin{abstract}
The concept of parity describes the inversion symmetry of a system and is of fundamental relevance in the standard model, quantum information processing, and field theory. In quantum electrodynamics, parity is conserved and large field gradients are required to engineer the parity of the light-matter interaction operator. In this work, we engineer a potassium-like artificial atom represented by a specifically designed superconducting flux qubit. We control the wave function parity of the artificial atom with an effective orbital momentum provided by a resonator. By irradiating the artificial atom with spatially shaped microwave fields, we select the interaction parity in situ. In this way, we observe dipole and quadrupole selection rules for single state transitions and induce transparency via longitudinal coupling. Our work advances the design of tunable artificial multilevel atoms to a new level, which is particularly promising with respect to quantum chemistry simulations with near-term superconducting circuits.
\end{abstract}

\maketitle

Parity and its underlying symmetries play an elementary role in pioneering theories such as CP violation~\cite{Wu_1957}, the Higgs formalism~\cite{Higgs_1964}, and quantum phase transitions~\cite{Trenkwalder_2016}. Parity measurements are essential in quantum information processing~\cite{Riste_2013,Corcoles_2015,Ofek_2016}, field theory~\cite{Felicetti_2015,Birrittella_2015}, and light-matter interaction~\cite{Liu_2005,Deppe_2008,Kou_2017}. In the latter, the application of high-frequency electromagnetic radiation in resonance with a desired state transition reveals the internal structure of matter. Depending on both symmetry and strength of the probe field, various multipole moments can be activated~\cite{Jain_2012} and the corresponding selection rules (SRs) become apparent~\cite{Rafac_2000}. In atomic systems, their natural properties limit systematic studies of SRs and often the interaction is dominated by the odd-parity dipole operator. Due to the small atomic length scale, even-parity quadrupole interactions are mostly relevant for large field gradients. Here, we therefore use a specific superconducting artificial atom~\cite{Paauw_2009,Schwarz_2013}, which is suitable for future experiments in quantum simulation and quantum chemistry~\cite{Mostame_2012,Mostame_2016}. In comparison to previous studies~\cite{Deppe_2008,Srinivasan_2011,You_2011,deGroot_2010,Didier_2015,Forn-Diaz_2015,Wu_2016,Royer_2017,Vool_2018}, our approach aims for the simulation of systems including orbital momentum by choosing a qubit architecture, where a single loop introduces a magnetic dipole moment and a gradiometer loop gives rise to a magnetic quadrupole moment. With those spatial degrees of freedom, we create an artificial potassium-like atom by introducing an effective orbital momentum provided by the wavefunction parity of a superconducting resonator. A carefully tailored pair of antennas additionally allows us to tune the parity of the light-matter interaction operator. Adjusting drive amplitude and phase, we can invert the RF magnetic field within the dimensions of the atom, which is a unique property in quantum optics. Changing the interaction parity is possible because the dipole moment provides longitudinal coupling~\cite{Billangeon_2015} (even interaction parity) and the quadrupole moment transversal coupling (odd interaction parity)~\cite{Liu_2005}. With this implementation we induce transparency~\cite{Liu_2014,Wu_2016} even when the potential of the artificial atom on its own does not exhibit a well-defined symmetry. Using the dispersively coupled resonator, we activate dipole forbidden transitions via sideband transitions~\cite{Blais_2007}. The ability to address dipolar and quadrupolar transitions in a single artificial atom together with a more complex level structure is of potential use in future quantum simulations of chemical compounds and quantum annealing.

\begin{figure}[t]
\includegraphics{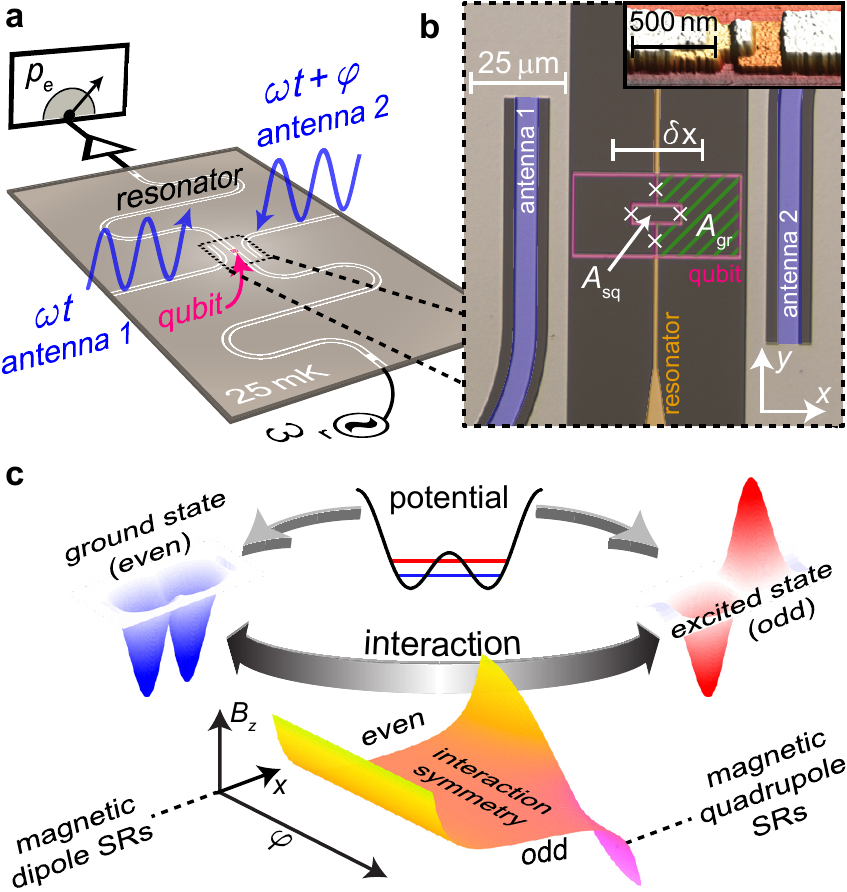}
\caption{\label{fig:Fig_12}\textbf{a} Chip layout and detection scheme. \textbf{b} False-colored micrograph of the qubit architecture. Crosses indicate Josephson junctions as the one shown in the atomic force micrograph (inset). The SQUID is placed on the symmetry axis of the qubit and the center point of the two gradiometer loops (green shaded area). \textbf{c} The symmetric double well potential for flux qubits results in two eigenstates with opposite parity (blue and red wave functions). The interaction is defined by the symmetry of a drive field that can be even, odd, or without a specific symmetry, controlled by the relative phase $\varphi$ between two frequency-degenerate microwave drives.}
\end{figure}

As shown in Fig.\,\ref{fig:Fig_12}a, we couple a tunable-gap gradiometric flux qubit~\cite{Paauw_2009,Schwarz_2013} galvanically to the fundamental current mode of a half-wavelength coplanar waveguide resonator operating at $\omega_{\mathrm{r}}/2\pi\,{\simeq}\,\SI{3.9}{\giga\hertz}$~\cite{supp}. In the first part of this work, we use the resonator only for readout purposes and in the second part, we additionally use the resonator to probe symmetry effects beyond the two-level atom approximation. The qubit itself consists of a gradiometric aluminum loop patterned on a high-resistivity silicon substrate (Fig.\,\ref{fig:Fig_12}b). The center conductor is interrupted by two Josephson junctions and a DC~SQUID where the latter provides an effective Josephson junction with a tunable critical current. To avoid interface losses~\cite{Goetz_2016}, the sample is fabricated from a single Al/AlO$_{\mathrm{x}}$/Al trilayer structure using the shadow evaporation technique~\cite{Dolan_1977}. We mount the sample to the \SI{25}{\milli\kelvin} base temperature stage of a custom-made dilution refrigerator. The resonator with Hamiltonian $\widehat{\mathcal{H}}_{\mathrm{r}}\,{=}\,\hbar\omega_{\mathrm{r}}\hat{a}^{\dag}\hat{a}$ is characterized by its external coupling rate $\kappa_{\mathrm{x}}/2\pi\,{\simeq}\,\SI{2.47}{\mega\hertz}$ and internal loss rate $\kappa_{\mathrm{i}}/2\pi\,{\simeq}\,\SI{70}{\kilo\hertz}$. The qubit excited state probability $p_{e}$ and the coherence times $T_{1}\,{\simeq}\,\SI{2.6}{\micro\second}$ and $T_{2}\,{\simeq}\,\SI{0.1}{\micro\second}$ are obtained with a dispersive readout scheme~\cite{Goetz_2016a}. The qubit-resonator coupling strength $g/2\pi\,{\simeq}\,\SI{41}{\mega\hertz}$ is predominantly transversal and is described by a light-matter interaction term $\widehat{\mathcal{H}}_{\mathrm{qr}}\,{=}\,\hbar(g_{\mathrm{t}}\cos\theta\hat{\sigma}_{z}\,{-}\,g_{\mathrm{t}}\sin\theta\hat{\sigma}_{x})(\hat{a}^{\dagger}+\hat{a})$. We trap one flux quantum in the outer qubit loop to generate a double well potential (Fig.\,\ref{fig:Fig_12}c). In terms of the odd- and even-parity Pauli operators $\hat{\sigma}_{x}$ and $\hat{\sigma}_{z}$, the resulting Hamiltonian reads $\widehat{\mathcal{H}}_{\mathrm{q}}^{\prime}\,{=}\,(\Delta\hat{\sigma}_{x}\,{+}\,\varepsilon\hat{\sigma}_{z})/2$. We set the tunnel coupling between the wells to $\Delta/\hbar\,{\simeq}\,2\pi\,{\times}\,\SI{8.2}{\giga\hertz}$ and vary the energy bias $\varepsilon$ with a pair of on-chip antennas placed symmetrically with respect to the qubit. The relevant $z$-component of the magnetic field irradiated by the antennas contains DC and AC components, i.e., $B_{z}(x,y)\,{=}\,B_{z}^{\mathrm{dc}}(x,y)\,{+}\,B_{z}^{\mathrm{ac}}(x,y)\cos\omega t$, where $\omega/2\pi$ is the drive frequency. Depending on the relative phase $\varphi$ between the two drive fields, we generate either a symmetric field configuration, $B_{z,\mathrm{sq}}\,{\equiv}\,\Phi_{\mathrm{sq}}/\mathcal{A}_{\mathrm{sq}}$ and $\delta B_{z,\mathrm{gr}}/\delta x\,{=}\,0$, or antisymmetric field gradients, $\delta B_{z,\mathrm{gr}}/\delta x\,{\equiv}\,(2\Phi_{\mathrm{gr}}/\mathcal{A}_{\mathrm{gr}})/\delta x$ and $B_{z,\mathrm{gr}}\,{=}\,0$. The fluxes $\Phi_{\mathrm{sq}}$ and $\Phi_{\mathrm{gr}}$ are the integrals of $B_{z}(x,y)$ over the areas $\mathcal{A}_{\mathrm{sq}}$ and $\mathcal{A}_{\mathrm{gr}}$ of the SQUID and a single gradiometer loop, respectively (see Fig.\,\ref{fig:Fig_12}b). For arbitrary $\varphi$, the total field $B_{z,\mathrm{tot}}^{\phantom{0}}\,{=}\,B_{z,\mathrm{sq}}\,{+}\,\delta B_{z,\mathrm{gr}}^{\phantom{0}}/\delta x$ is a superposition of symmetric and antisymmetric contributions.

Due to the elaborate sample geometry, the two terms of $B_{z,\mathrm{tot}}$ couple to different Pauli operators in the Hamiltonian $\widehat{\mathcal{H}}\,{=}\,pB_{z,\mathrm{sq}}\hat{\sigma}_{z}\,{+}\,Q(\delta B_{z,\mathrm{gr}}/\delta x)\hat{\sigma}_{x}$~\cite{supp}. Here, the SQUID dipole moment $p$ and the gradiometer quadrupole moment $Q$ define the longitudinal and transversal coupling strengths $\Omega_{\ell}(\varphi)\,{=}\,p B_{z,\mathrm{sq}}^{\mathrm{ac}}(\varphi)/\hbar$ and $\Omega_{\mathrm{t}}(\varphi)\,{=}\,Q(\delta B_{z,\mathrm{gr}}^{\mathrm{ac}}(\varphi)/\delta x)/\hbar$, respectively. Hence, the qubit couples longitudinally to symmetric fields $(\varphi\,{\in}\,\{0,2\pi,...\}\,{\rightarrow}\,\Omega_{\mathrm{t}}\,{=}\,0)$ and transversally to antisymmetric fields $(\varphi\,{\in}\,\{\pi,3\pi,...\}\,{\rightarrow}\,\Omega_{\ell}\,{=}\,0)$. In the qubit energy eigenbasis, where $\widehat{\mathcal{H}}_{\mathrm{q}}\,{=}\,\hbar\omega_{\mathrm{q}}\hat{\sigma}_{z}/2\,{\equiv}\,\sqrt{\Delta^{2}\,{+}\,\varepsilon^{2}}\hat{\sigma}_{z}/2$ and $\theta\,{=}\,\tan^{-1}(\Delta/\varepsilon)$ is the Bloch angle controlled by $B_{z}^{\mathrm{dc}}$, the interaction Hamiltonian reads

\begin{align}
\widehat{\mathcal{H}}_{\mathrm{int}} =  \hbar \cos(\omega t)\big[\big(\Omega_{\ell}(\varphi)\cos\theta &- \Omega_{\mathrm{t}}(\varphi)\sin\theta\big)\hat{\sigma}_{x}\notag\\
 + \big(\Omega_{\mathrm{t}}(\varphi)\cos\theta &+ \Omega_{\ell}(\varphi)\sin\theta\big)\hat{\sigma}_{z}\big]/2\,.
\label{eqn:Hqd}
\end{align}

In addition to the field symmetry characterized by $\Omega_{\ell}$ and $\Omega_{\mathrm{t}}$ as described above, the parity of $\widehat{\mathcal{H}}_{\mathrm{int}}$ depends on the symmetry of the qubit wave functions. The latter is governed by the qubit parity operator $\widehat{\Pi}_{\mathrm{q}}\,{=}\,{-}\hat{\sigma}_{z}$, which results in an odd parity for pure $\hat{\sigma}_{x}$ interaction, an even parity for pure $\hat{\sigma}_{z}$ interaction, and no well-defined parity when both terms are present~\cite{Deppe_2008}. Even interaction parity results in forbidden transitions (transparency) because the commutator $[\hat{\Pi}_{\mathrm{q}},\hat{\sigma}_{z}]\,{=}\,0$. Then, $\hat{\sigma}_{z}$ is a constant of motion, $\imath\hbar\partial\hat{\sigma}_{z}/\partial t\,{=}\,0$, and the qubit remains in the ground state with a modulated qubit gap $\Delta$. Since, in contrast to Ref.~\onlinecite{Deppe_2008}, we activate $\hat{\sigma}_{x}$ with an antisymmetric drive field at $\theta\,{=}\,\pi/2$, the most natural interpretation of our system is given in terms of magnetic SRs.

\begin{figure}[t]
\centering
\includegraphics{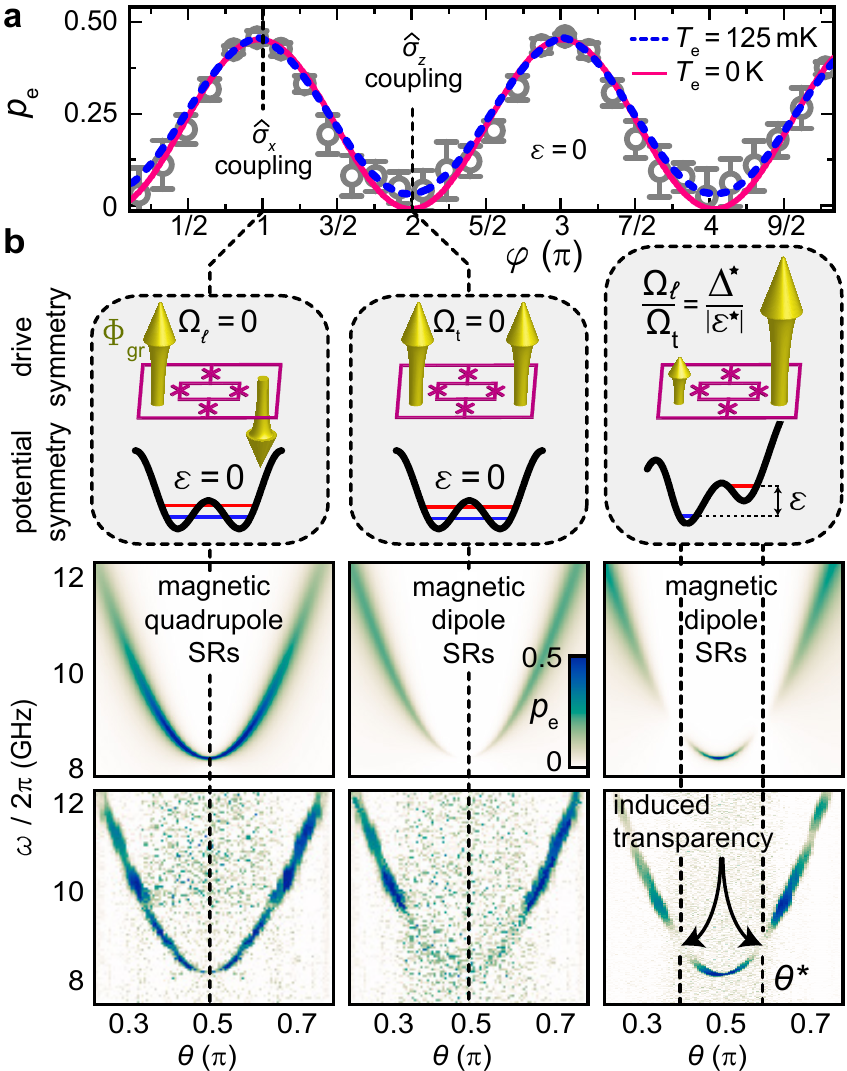}
\caption{\textbf{a} Qubit excited state probability $p_{\mathrm{e}}$ plotted versus relative phase $\varphi$ at $\theta\,{=}\,\pi/2$. The solid line is a $\sin^{2}(\varphi/2)$ fit assuming the qubit to be at zero temperature, the dashed line takes a finite qubit temperature into account and the error bars are of statistical nature. \textbf{b} Top panel: Symmetry of the drive field amplitude and the qubit potential. Lower panel: Simulated (top row) and measured (bottom row) excited state probability $p_{\mathrm{e}}$ plotted versus the Bloch angle $\theta$ and the drive frequency $\omega/2\pi$. At $\theta^{\star}$ we observe longitudinal coupling-induced transparency. We attribute the scatter between $0.3\pi\,{\lesssim}\,\theta\,{\lesssim}\,0.6\pi$ to variations in the signal-to-noise ratio of our measurement setup. The reduced signal strength at $\theta\,{\simeq}\,0.7$ in the bottom right panel is not predicted by our model and could be due to a stray microwave mode or a two-level defect coupled to the qubit.}
\label{fig:Fig_03}
\end{figure}

In a first experiment, we investigate the two-antenna control of the drive symmetry required to observe magnetic SRs. We operate at the point of optimal phase coherence, $\theta\,{=}\,\pi/2$, where the qubit potential is symmetric and initial state $\ket{\mathrm{g}}$ and final $\ket{\mathrm{e}}$ state have opposite parity~\cite{Liu_2005}, $\Pi_{\mathrm{i}}\,{=}\,{-}\Pi_{\mathrm{f}}$. In Fig.\,\ref{fig:Fig_03}a, we plot the qubit excited state probability $p_{\mathrm{e}}\,{=}\,(\langle\hat{\sigma}_{z}\rangle\,{+}\,1)/2$ for different spatial distributions controlled by the relative phase $\varphi$. The resonant drive $(\omega\,{=}\,\omega_{\mathrm{q}})$ is kept at constant power and is split symmetrically into the two antenna lines~\cite{supp}. We observe oscillations between $p_{\mathrm{e}}\,{\simeq}\,0$ and $p_{\mathrm{e}}\,{\simeq}\,0.5$ occurring when $\varphi$ assumes integer multiples of $\pi$. For these values, also the drive has a well-defined parity which leads to the formation of magnetic SRs. As long as $\varphi$ is not set to an integer multiple of $\pi$, we shape a drive field without specific symmetry $(\Omega_{\mathrm{t}},\Omega_{\ell}\,{\neq}\,0)$. Then, there are no SRs and $p_{\mathrm{e}}$ follows qualitatively the expected $\sin^{2}(\varphi/2)$-dependence. There is a small stray excitation probability $p_{\mathrm{e}}^{\mathrm{str}}\,{\simeq}\,0.05\,{=}\,\exp[-\hbar\omega_{\mathrm{q}}/(k_{\mathrm{B}}T_{\mathrm{e}})]$ for pure $\hat{\sigma}_{z}$ interaction, which can be linked to an effective temperature $T_{\mathrm{e}}$. We attribute this effect to a constant thermal contribution in the transversal drive, $\hbar\Omega_{\mathrm{t}}\,{\mapsto}\,\hbar\Omega_{\mathrm{t}}(\varphi)\,{+}\,p_{\mathrm{e}}^{\mathrm{str}}\hbar\omega_{\mathrm{q}}$. Using this ansatz, we find quantitative agreement between theory and experiment for $T_{\mathrm{e}}\,{\simeq}\,\SI{125}{\milli\kelvin}$. Similar values are found in other experiments on superconducting circuits~\cite{Yan_2015,Goetz_2016a,Tan_2017}.
 
In the next step, we demonstrate an enhanced level of control by investigating SRs for a tilted qubit potential, $\theta\,{\neq}\,\pi/2$. To this end, we measure the averaged excited state probability $p_{e}$ as a function of drive frequency $\omega$ and potential tilt, which is controlled by $\theta$. In the cases of purely symmetric and antisymmetric drives, which are shown in the left and middle column of Fig.\,\ref{fig:Fig_03}b, the broken symmetry of the qubit potential leads to the absence of SRs~\cite{Deppe_2008}. Clearly, the transition is allowed for both symmetric and antisymmetric drive fields away from the optimal point. Our specific qubit and antenna geometry allows us to restore the broken symmetry and observe rigorous SRs also for a tilted qubit potential. To this end, we sweep $\theta$ for a certain ratio $\Omega_{\ell}/\Omega_{\mathrm{t}}\,{\simeq}\,30$ and observe a strong decrease in $p_{\mathrm{e}}$ for $\theta^{\star}\,{\simeq}\,0.4\pi$ and $\theta^{\star}\,{\simeq}\,0.6\pi$ as shown in the right column of Fig.\,\ref{fig:Fig_03}b. At these points, the ratio $\Omega_{\ell}/\Omega_{\mathrm{t}}$ is equal to the asymmetry parameter $\tan\theta^{\star}\,{\equiv}\,\Delta^{\star}/|\varepsilon^{\star}|$ of the qubit potential. In other words, we exploit that the presence of SRs is determined by the symmetry properties of the interaction operator $\widehat{\mathcal{H}}_{\mathrm{int}}$, rather than by those of the qubit alone. The vanishing transition matrix element at $\theta^{\star}$ is known as longitudinal coupling-induced transparency~\cite{Liu_2014}. It can be understood from a formal point of view, because the $\hat{\sigma}_{x}$-term in the Hamiltonian of Eq.\,(\ref{eqn:Hqd}) vanishes when $\Omega_{\ell}/\Omega_{\mathrm{t}}\,{=}\,\Delta^{\star}/|\varepsilon^{\star}|$. In this case, one can rotate into a basis where both qubit potential and drive field are symmetric such that the interaction Hamiltonian becomes $\widehat{\mathcal{H}}_{\mathrm{int}}^{\star}\,{=}\,\Omega_{\mathrm{t}}\hat{\sigma}_{z}/\cos\theta^{\star}$.

\begin{figure*}[t!]
\centering
\includegraphics{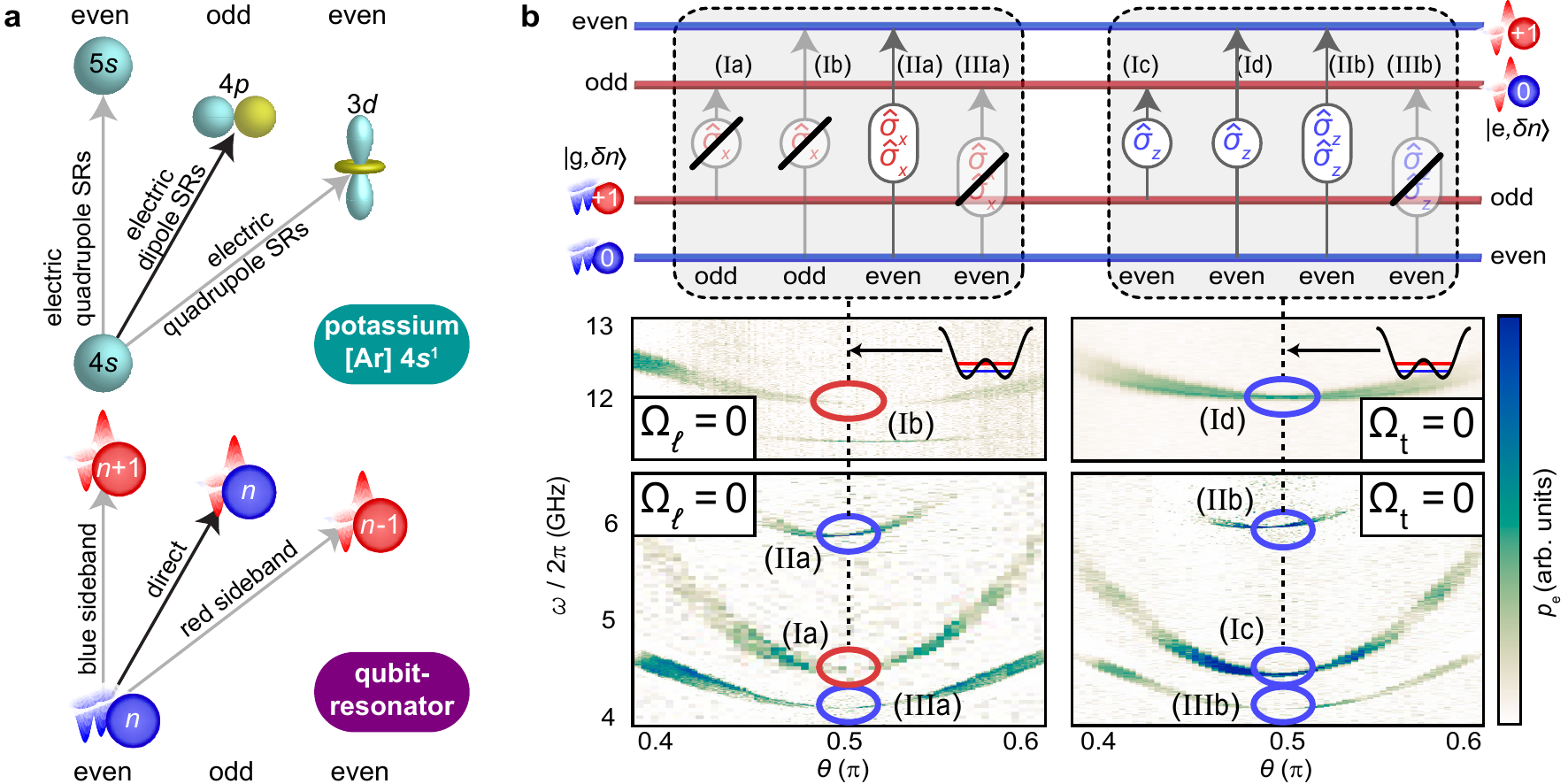}
\caption{\textbf{a} Energy level diagram for ground and first three excited states for a potassium atom (top) and the qubit-resonator system (bottom). The atom follows electric SRs, which have a close analogy to sideband transitions in the qubit-resonator system. \textbf{b} Top panel: Level scheme and corresponding parity of composite qubit-resonator states for multi-photon transitions $\ket{\mathrm{g},\delta n}\,{\mapsto}\,\ket{\mathrm{e},{\pm}1}$. Crossed out arrows denote forbidden transitions. Lower panel: Qubit excited state probability $p_{\mathrm{e}}$ plotted versus Bloch angle $\theta$ and drive frequency $\omega/2\pi$. The circled areas show the red and blue sidebands (I\,a-d), the two-photon transition of the blue sideband (II\,a,b), and the direct two-photon transition (III\,a,b) for a symmetric qubit potential. Due to different power levels required to drive the transitions, the data is an overlay of different measurements. The vertical shift between the cases II\,a and II\,b is caused by small flux jumps between the measurements.}
\label{fig:Fig_04}
\end{figure*}

So far, our discussion has been restricted to an artificial two-level atom, where the qubit states $\ket{\mathrm{g}}$ and $\ket{\mathrm{e}}$ behave similar to the $1s$ and $2p$ state of a hydrogen atom, respectively, and transparency due to longitudinal coupling could be induced. To observe also the opposite phenomenon, the activation of transversally forbidden transitions between states of equal parity, we engineer a more complex artificial atom using the dispersively coupled resonator. In the parity operator for the resonator,~\cite{Ofek_2016} $\widehat{\Pi}_{\mathrm{r}}\,{=}\,e^{\imath\pi n}$, the eigenvalues $n$ of the photon number operator $\hat{n}$ play the role of the quantum number of an orbital momentum in natural atoms. As shown in Fig.\,\ref{fig:Fig_04}a, we now exploit the parity~\cite{Birrittella_2015} $\widehat{\Pi}_{\mathrm{q}}\widehat{\Pi}_{\mathrm{r}}\,{=}\,{-}\hat{\sigma}_{z}e^{\imath\pi n}$ of the combined qubit-resonator system and map the state $\ket{\mathrm{g},n}$ to the even $s$-orbital. Similarly, when the qubit is in the excited state, the resonator states $\ket{n\,{+}\,1}$, $\ket{n}$, and $\ket{n\,{-}\,1}$ give rise to artificial even $s$-, odd $p$-, and even $d$-orbitals, respectively. We note that, although the qubit parity enters into the details of our mapping, it is essentially the resonator, which acts as an artificial orbital momentum. In this way, the qubit-resonator system qualitatively mimics the energy levels and parity properties of a potassium atom with electron configuration $[\mathrm{Ar}]4s^{1}$ (see Fig.\,\ref{fig:Fig_04}a). The corresponding electric SRs are precisely reproduced if we realize the symmetric (antisymmetric) electric drive as a physical antisymmetric (symmetric) magnetic field. In order to avoid confusion with our previous discussion of magnetic SRs, we prefer to present the experimental results in the more universal language of transversal and longitudinal coupling. We first investigate the effect of SRs between states of equal parity by probing the red and blue sideband transitions, $\ket{\mathrm{g},n}\,{\mapsto}\,\ket{\mathrm{e},n{\pm}1}$ (cases Ia\,{--}\,Id in Fig.\,\ref{fig:Fig_04}b). To this end, we probe the resonator response at frequency $\omega_{\mathrm{r}}$ while simultaneously driving the sidebands at frequencies $\omega\,{=}\,\omega_{\mathrm{q}}\,{\pm}\,\omega_{\mathrm{r}}$. The photon required for the absorption process of the red sideband transition, $\ket{\mathrm{g},n}\,{\mapsto}\,\ket{\mathrm{e},n{-}1}$, is provided by the readout tone, which contains approximately $\langle n\rangle\,{\simeq}\,30$ photons. Without longitudinal coupling ($\Omega_{\ell}\,{=}\,0$), resonant transitions between states of equal parity are forbidden and both sidebands vanish at $\theta\,{=}\,\pi/2$. Changing the relative phase $\varphi$ to apply a purely longitudinal drive $(\Omega_{\mathrm{t}}\,{=}\,0)$, these transitions become allowed and we observe a strong increase in $p_{\mathrm{e}}$ at $\theta\,{=}\,\pi/2$. Interestingly, the transition is always allowed for the two-photon blue sideband [case (IIa) and case (IIb)] and always forbidden for the two-photon qubit excitation [case (IIIa) and case (IIIb) in Fig.\,\ref{fig:Fig_04}b]. The reason for this behavior is that applying an odd or an even operator twice always produces an even-parity operator~\cite{Blais_2007}. Despite the more complex scenario, all transitions are allowed when the symmetry of the qubit potential or the interaction is not purely transversal or purely longitudinal. We note that higher-order sideband transitions can be neglected in our experiments because they are detuned strongly in frequency and typically require a much higher driving power to be activated.

In conclusion, we have established a superconducting architecture allowing for full in situ control over the parity of light-matter interaction and the participating quantum states. We have selectively addressed the magnetic dipole and quadrupole moment of our artificial atom and verified the corresponding SRs. Furthermore, we show that a dispersively coupled resonator acts as an artificial orbital momentum and enables the simulation of multilevel artificial atoms. Even though the potassium level structure could, in principle, be obtained using other qubit types, we find that the precise shaping of interaction symmetries works particularly well with the gradiometric tunable-gap flux qubit. This central result will be essential for future simulations of quantum chemistry~\cite{Mostame_2012,Mostame_2016,Potocnik_2018}. Especially regarding near-term quantum devices with a relatively small number of quantum circuits, the prospect of obtaining a higher complexity from replacing qubits with more realistic artificial multilevel atoms is very appealing. In the system discussed here, even spin-orbit coupling could be realized by adding resonators with tunable qubit-resonator coupling strengths. More generally speaking, the coexistence of longitudinal and transversal coupling is important to reach the non-stoquastic regime in adiabatic quantum computing~\cite{Hormozi_2017,Albash_2018} and quantum annealing~\cite{Brooke_1999,Dorit_2007,Lanting_2014,Vinci_2017}. Based on our results on highly tunable qubit-drive coupling, the controlled coexistence of transversal and longitudinal interactions between two gradiometric tunable-gap qubits appears feasible by designing their magnetic field distribution, by coupling the qubit to quantum circuits providing symmetric and antisymmetric modes~\cite{Baust_2015,Baust_2016}, or using a parametric drive. In addition, for increasing coupling strengths, longitudinal coupling terms may lead to new experimental studies of the symmetries in the quantum Rabi model~\cite{Braak_2011,Felicetti_2015} and help to obtain the quantum Fisher information for probabilistic parameters in time-dependent Hamiltonians~\cite{Pang_2017}. The range of possible applications can be even further extended by considering longitudinal qubit-resonator coupling~\cite{Richer_2016}, quantum state engineering~\cite{Porras_2012,Quijandria_2013,Zhao_2015}, and the simulation of relativistic physics~\cite{delRey_2012}.

\begin{acknowledgments}
We thank E.~Solano and D.~Braak for fruitful discussions. We acknowledge financial support from the German Research Foundation through SFB 631 and FE 1564/1-1, the doctorate program ExQM of the Elite Network of Bavaria, and the International Max Planck Research School "Quantum Science and Technology``. The authors declare no competing financial interests.
\end{acknowledgments}

\bibliography{D:/Dropbox/Goetz_Bibliography}

\begin{thebibliography}{61}%
\makeatletter
\providecommand \@ifxundefined [1]{%
 \@ifx{#1\undefined}
}%
\providecommand \@ifnum [1]{%
 \ifnum #1\expandafter \@firstoftwo
 \else \expandafter \@secondoftwo
 \fi
}%
\providecommand \@ifx [1]{%
 \ifx #1\expandafter \@firstoftwo
 \else \expandafter \@secondoftwo
 \fi
}%
\providecommand \natexlab [1]{#1}%
\providecommand \enquote  [1]{``#1''}%
\providecommand \bibnamefont  [1]{#1}%
\providecommand \bibfnamefont [1]{#1}%
\providecommand \citenamefont [1]{#1}%
\providecommand \href@noop [0]{\@secondoftwo}%
\providecommand \href [0]{\begingroup \@sanitize@url \@href}%
\providecommand \@href[1]{\@@startlink{#1}\@@href}%
\providecommand \@@href[1]{\endgroup#1\@@endlink}%
\providecommand \@sanitize@url [0]{\catcode `\\12\catcode `\$12\catcode
  `\&12\catcode `\#12\catcode `\^12\catcode `\_12\catcode `\%12\relax}%
\providecommand \@@startlink[1]{}%
\providecommand \@@endlink[0]{}%
\providecommand \url  [0]{\begingroup\@sanitize@url \@url }%
\providecommand \@url [1]{\endgroup\@href {#1}{\urlprefix }}%
\providecommand \urlprefix  [0]{URL }%
\providecommand \Eprint [0]{\href }%
\providecommand \doibase [0]{http://dx.doi.org/}%
\providecommand \selectlanguage [0]{\@gobble}%
\providecommand \bibinfo  [0]{\@secondoftwo}%
\providecommand \bibfield  [0]{\@secondoftwo}%
\providecommand \translation [1]{[#1]}%
\providecommand \BibitemOpen [0]{}%
\providecommand \bibitemStop [0]{}%
\providecommand \bibitemNoStop [0]{.\EOS\space}%
\providecommand \EOS [0]{\spacefactor3000\relax}%
\providecommand \BibitemShut  [1]{\csname bibitem#1\endcsname}%
\let\auto@bib@innerbib\@empty
%</preamble>
\bibitem [{\citenamefont {Wu}\ \emph {et~al.}(1957)\citenamefont {Wu},
  \citenamefont {Ambler}, \citenamefont {Hayward}, \citenamefont {Hoppes},\
  and\ \citenamefont {Hudson}}]{Wu_1957}%
  \BibitemOpen
  \bibfield  {author} {\bibinfo {author} {\bibfnamefont {C.~S.}\ \bibnamefont
  {Wu}}, \bibinfo {author} {\bibfnamefont {E.}~\bibnamefont {Ambler}}, \bibinfo
  {author} {\bibfnamefont {R.~W.}\ \bibnamefont {Hayward}}, \bibinfo {author}
  {\bibfnamefont {D.~D.}\ \bibnamefont {Hoppes}}, \ and\ \bibinfo {author}
  {\bibfnamefont {R.~P.}\ \bibnamefont {Hudson}},\ }\href {\doibase
  10.1103/PhysRev.105.1413} {\bibfield  {journal} {\bibinfo  {journal}
  {Phys.\,Rev.}\ }\textbf {\bibinfo {volume} {105}},\ \bibinfo {pages} {1413}
  (\bibinfo {year} {1957})}\BibitemShut {NoStop}%
\bibitem [{\citenamefont {Higgs}(1964)}]{Higgs_1964}%
  \BibitemOpen
  \bibfield  {author} {\bibinfo {author} {\bibfnamefont {P.~W.}\ \bibnamefont
  {Higgs}},\ }\href {\doibase 10.1103/PhysRevLett.13.508} {\bibfield  {journal}
  {\bibinfo  {journal} {Phys.\,Rev.\,Lett.}\ }\textbf {\bibinfo {volume}
  {13}},\ \bibinfo {pages} {508} (\bibinfo {year} {1964})}\BibitemShut
  {NoStop}%
\bibitem [{\citenamefont {Trenkwalder}\ \emph {et~al.}(2016)\citenamefont
  {Trenkwalder}, \citenamefont {Spagnolli}, \citenamefont {Semeghini},
  \citenamefont {Coop}, \citenamefont {Landini}, \citenamefont {Castilho},
  \citenamefont {Pezze}, \citenamefont {Modugno}, \citenamefont {Inguscio},
  \citenamefont {Smerzi},\ and\ \citenamefont {Fattori}}]{Trenkwalder_2016}%
  \BibitemOpen
  \bibfield  {author} {\bibinfo {author} {\bibfnamefont {A.}~\bibnamefont
  {Trenkwalder}}, \bibinfo {author} {\bibfnamefont {G.}~\bibnamefont
  {Spagnolli}}, \bibinfo {author} {\bibfnamefont {G.}~\bibnamefont
  {Semeghini}}, \bibinfo {author} {\bibfnamefont {S.}~\bibnamefont {Coop}},
  \bibinfo {author} {\bibfnamefont {M.}~\bibnamefont {Landini}}, \bibinfo
  {author} {\bibfnamefont {P.}~\bibnamefont {Castilho}}, \bibinfo {author}
  {\bibfnamefont {L.}~\bibnamefont {Pezze}}, \bibinfo {author} {\bibfnamefont
  {G.}~\bibnamefont {Modugno}}, \bibinfo {author} {\bibfnamefont
  {M.}~\bibnamefont {Inguscio}}, \bibinfo {author} {\bibfnamefont
  {A.}~\bibnamefont {Smerzi}}, \ and\ \bibinfo {author} {\bibfnamefont
  {M.}~\bibnamefont {Fattori}},\ }\href {\doibase 10.1038/nphys3743} {\bibfield
   {journal} {\bibinfo  {journal} {Nat.\,Phys.}\ }\textbf {\bibinfo {volume}
  {12}},\ \bibinfo {pages} {826} (\bibinfo {year} {2016})}\BibitemShut
  {NoStop}%
\bibitem [{\citenamefont {Riste}\ \emph {et~al.}(2013)\citenamefont {Riste},
  \citenamefont {Bultink}, \citenamefont {Tiggelman}, \citenamefont {Schouten},
  \citenamefont {Lehnert},\ and\ \citenamefont {DiCarlo}}]{Riste_2013}%
  \BibitemOpen
  \bibfield  {author} {\bibinfo {author} {\bibfnamefont {D.}~\bibnamefont
  {Riste}}, \bibinfo {author} {\bibfnamefont {C.~C.}\ \bibnamefont {Bultink}},
  \bibinfo {author} {\bibfnamefont {M.~J.}\ \bibnamefont {Tiggelman}}, \bibinfo
  {author} {\bibfnamefont {R.~N.}\ \bibnamefont {Schouten}}, \bibinfo {author}
  {\bibfnamefont {K.~W.}\ \bibnamefont {Lehnert}}, \ and\ \bibinfo {author}
  {\bibfnamefont {L.}~\bibnamefont {DiCarlo}},\ }\href {\doibase
  10.1038/ncomms2936} {\bibfield  {journal} {\bibinfo  {journal} {Nat.\,Comm.}\
  }\textbf {\bibinfo {volume} {4}},\ \bibinfo {pages} {1913} (\bibinfo {year}
  {2013})}\BibitemShut {NoStop}%
\bibitem [{\citenamefont {C{\'o}rcoles}\ \emph {et~al.}(2015)\citenamefont
  {C{\'o}rcoles}, \citenamefont {Magesan}, \citenamefont {Srinivasan},
  \citenamefont {Cross}, \citenamefont {Steffen}, \citenamefont {Gambetta},\
  and\ \citenamefont {Chow}}]{Corcoles_2015}%
  \BibitemOpen
  \bibfield  {author} {\bibinfo {author} {\bibfnamefont {A.}~\bibnamefont
  {C{\'o}rcoles}}, \bibinfo {author} {\bibfnamefont {E.}~\bibnamefont
  {Magesan}}, \bibinfo {author} {\bibfnamefont {S.~J.}\ \bibnamefont
  {Srinivasan}}, \bibinfo {author} {\bibfnamefont {A.~W.}\ \bibnamefont
  {Cross}}, \bibinfo {author} {\bibfnamefont {M.}~\bibnamefont {Steffen}},
  \bibinfo {author} {\bibfnamefont {J.~M.}\ \bibnamefont {Gambetta}}, \ and\
  \bibinfo {author} {\bibfnamefont {J.~M.}\ \bibnamefont {Chow}},\ }\href
  {\doibase 10.1038/ncomms7979} {\bibfield  {journal} {\bibinfo  {journal}
  {Nat.\,Comm.}\ }\textbf {\bibinfo {volume} {6}},\ \bibinfo {pages} {6979}
  (\bibinfo {year} {2015})}\BibitemShut {NoStop}%
\bibitem [{\citenamefont {Ofek}\ \emph {et~al.}(2016)\citenamefont {Ofek},
  \citenamefont {Petrenko}, \citenamefont {Heeres}, \citenamefont {Reinhold},
  \citenamefont {Leghtas}, \citenamefont {Vlastakis}, \citenamefont {Liu},
  \citenamefont {Frunzio}, \citenamefont {Girvin}, \citenamefont {Jiang},
  \citenamefont {Mirrahimi}, \citenamefont {Devoret},\ and\ \citenamefont
  {Schoelkopf}}]{Ofek_2016}%
  \BibitemOpen
  \bibfield  {author} {\bibinfo {author} {\bibfnamefont {N.}~\bibnamefont
  {Ofek}}, \bibinfo {author} {\bibfnamefont {A.}~\bibnamefont {Petrenko}},
  \bibinfo {author} {\bibfnamefont {R.}~\bibnamefont {Heeres}}, \bibinfo
  {author} {\bibfnamefont {P.}~\bibnamefont {Reinhold}}, \bibinfo {author}
  {\bibfnamefont {Z.}~\bibnamefont {Leghtas}}, \bibinfo {author} {\bibfnamefont
  {B.}~\bibnamefont {Vlastakis}}, \bibinfo {author} {\bibfnamefont
  {Y.}~\bibnamefont {Liu}}, \bibinfo {author} {\bibfnamefont {L.}~\bibnamefont
  {Frunzio}}, \bibinfo {author} {\bibfnamefont {S.~M.}\ \bibnamefont {Girvin}},
  \bibinfo {author} {\bibfnamefont {L.}~\bibnamefont {Jiang}}, \bibinfo
  {author} {\bibfnamefont {M.}~\bibnamefont {Mirrahimi}}, \bibinfo {author}
  {\bibfnamefont {M.~H.}\ \bibnamefont {Devoret}}, \ and\ \bibinfo {author}
  {\bibfnamefont {R.~J.}\ \bibnamefont {Schoelkopf}},\ }\href {\doibase
  10.1038/nature18949} {\bibfield  {journal} {\bibinfo  {journal} {Nature}\
  }\textbf {\bibinfo {volume} {536}},\ \bibinfo {pages} {441} (\bibinfo {year}
  {2016})}\BibitemShut {NoStop}%
\bibitem [{\citenamefont {Felicetti}\ \emph {et~al.}(2015)\citenamefont
  {Felicetti}, \citenamefont {Douce}, \citenamefont {Romero}, \citenamefont
  {Milman},\ and\ \citenamefont {Solano}}]{Felicetti_2015}%
  \BibitemOpen
  \bibfield  {author} {\bibinfo {author} {\bibfnamefont {S.}~\bibnamefont
  {Felicetti}}, \bibinfo {author} {\bibfnamefont {T.}~\bibnamefont {Douce}},
  \bibinfo {author} {\bibfnamefont {G.}~\bibnamefont {Romero}}, \bibinfo
  {author} {\bibfnamefont {P.}~\bibnamefont {Milman}}, \ and\ \bibinfo {author}
  {\bibfnamefont {E.}~\bibnamefont {Solano}},\ }\href {\doibase
  10.1038/srep11818} {\bibfield  {journal} {\bibinfo  {journal} {Sci.\,Rep.}\
  }\textbf {\bibinfo {volume} {5}},\ \bibinfo {pages} {11818} (\bibinfo {year}
  {2015})}\BibitemShut {NoStop}%
\bibitem [{\citenamefont {Birrittella}\ \emph {et~al.}(2015)\citenamefont
  {Birrittella}, \citenamefont {Cheng},\ and\ \citenamefont
  {Gerry}}]{Birrittella_2015}%
  \BibitemOpen
  \bibfield  {author} {\bibinfo {author} {\bibfnamefont {R.}~\bibnamefont
  {Birrittella}}, \bibinfo {author} {\bibfnamefont {K.}~\bibnamefont {Cheng}},
  \ and\ \bibinfo {author} {\bibfnamefont {C.~C.}\ \bibnamefont {Gerry}},\
  }\href {\doibase 10.1016/j.optcom.2015.05.069} {\bibfield  {journal}
  {\bibinfo  {journal} {Opt. Commun.}\ }\textbf {\bibinfo {volume} {354}},\
  \bibinfo {pages} {286} (\bibinfo {year} {2015})}\BibitemShut {NoStop}%
\bibitem [{\citenamefont {Liu}\ \emph {et~al.}(2005)\citenamefont {Liu},
  \citenamefont {You}, \citenamefont {Wei}, \citenamefont {Sun},\ and\
  \citenamefont {Nori}}]{Liu_2005}%
  \BibitemOpen
  \bibfield  {author} {\bibinfo {author} {\bibfnamefont {Y.}~\bibnamefont
  {Liu}}, \bibinfo {author} {\bibfnamefont {J.~Q.}\ \bibnamefont {You}},
  \bibinfo {author} {\bibfnamefont {L.~F.}\ \bibnamefont {Wei}}, \bibinfo
  {author} {\bibfnamefont {C.~P.}\ \bibnamefont {Sun}}, \ and\ \bibinfo
  {author} {\bibfnamefont {F.}~\bibnamefont {Nori}},\ }\href {\doibase
  10.1103/PhysRevLett.95.087001} {\bibfield  {journal} {\bibinfo  {journal}
  {Phys.\,Rev.\,Lett.}\ }\textbf {\bibinfo {volume} {95}},\ \bibinfo {pages}
  {087001} (\bibinfo {year} {2005})}\BibitemShut {NoStop}%
\bibitem [{\citenamefont {Deppe}\ \emph {et~al.}(2008)\citenamefont {Deppe},
  \citenamefont {Mariantoni}, \citenamefont {Menzel}, \citenamefont {Marx},
  \citenamefont {Saito}, \citenamefont {Kakuyanagi}, \citenamefont {Tanaka},
  \citenamefont {Meno}, \citenamefont {Semba}, \citenamefont {Takayanagi},
  \citenamefont {Solano},\ and\ \citenamefont {Gross}}]{Deppe_2008}%
  \BibitemOpen
  \bibfield  {author} {\bibinfo {author} {\bibfnamefont {F.}~\bibnamefont
  {Deppe}}, \bibinfo {author} {\bibfnamefont {M.}~\bibnamefont {Mariantoni}},
  \bibinfo {author} {\bibfnamefont {E.~P.}\ \bibnamefont {Menzel}}, \bibinfo
  {author} {\bibfnamefont {A.}~\bibnamefont {Marx}}, \bibinfo {author}
  {\bibfnamefont {S.}~\bibnamefont {Saito}}, \bibinfo {author} {\bibfnamefont
  {K.}~\bibnamefont {Kakuyanagi}}, \bibinfo {author} {\bibfnamefont
  {H.}~\bibnamefont {Tanaka}}, \bibinfo {author} {\bibfnamefont
  {T.}~\bibnamefont {Meno}}, \bibinfo {author} {\bibfnamefont {K.}~\bibnamefont
  {Semba}}, \bibinfo {author} {\bibfnamefont {H.}~\bibnamefont {Takayanagi}},
  \bibinfo {author} {\bibfnamefont {E.}~\bibnamefont {Solano}}, \ and\ \bibinfo
  {author} {\bibfnamefont {R.}~\bibnamefont {Gross}},\ }\href {\doibase
  10.1038/nphys1016} {\bibfield  {journal} {\bibinfo  {journal} {Nat.\,Phys.}\
  }\textbf {\bibinfo {volume} {4}},\ \bibinfo {pages} {686} (\bibinfo {year}
  {2008})}\BibitemShut {NoStop}%
\bibitem [{\citenamefont {Kou}\ \emph {et~al.}(2017)\citenamefont {Kou},
  \citenamefont {Smith}, \citenamefont {Vool}, \citenamefont {Brierley},
  \citenamefont {Meier}, \citenamefont {Frunzio}, \citenamefont {Girvin},
  \citenamefont {Glazman},\ and\ \citenamefont {Devoret}}]{Kou_2017}%
  \BibitemOpen
  \bibfield  {author} {\bibinfo {author} {\bibfnamefont {A.}~\bibnamefont
  {Kou}}, \bibinfo {author} {\bibfnamefont {W.~C.}\ \bibnamefont {Smith}},
  \bibinfo {author} {\bibfnamefont {U.}~\bibnamefont {Vool}}, \bibinfo {author}
  {\bibfnamefont {R.~T.}\ \bibnamefont {Brierley}}, \bibinfo {author}
  {\bibfnamefont {H.}~\bibnamefont {Meier}}, \bibinfo {author} {\bibfnamefont
  {L.}~\bibnamefont {Frunzio}}, \bibinfo {author} {\bibfnamefont {S.~M.}\
  \bibnamefont {Girvin}}, \bibinfo {author} {\bibfnamefont {L.~I.}\
  \bibnamefont {Glazman}}, \ and\ \bibinfo {author} {\bibfnamefont {M.~H.}\
  \bibnamefont {Devoret}},\ }\href {\doibase 10.1103/PhysRevX.7.031037}
  {\bibfield  {journal} {\bibinfo  {journal} {Phys. Rev. X}\ }\textbf {\bibinfo
  {volume} {7}},\ \bibinfo {pages} {031037} (\bibinfo {year}
  {2017})}\BibitemShut {NoStop}%
\bibitem [{\citenamefont {Jain}\ \emph {et~al.}(2012)\citenamefont {Jain},
  \citenamefont {Ghosh}, \citenamefont {Baer}, \citenamefont {Rabani},\ and\
  \citenamefont {Alivisatos}}]{Jain_2012}%
  \BibitemOpen
  \bibfield  {author} {\bibinfo {author} {\bibfnamefont {P.~K.}\ \bibnamefont
  {Jain}}, \bibinfo {author} {\bibfnamefont {D.}~\bibnamefont {Ghosh}},
  \bibinfo {author} {\bibfnamefont {R.}~\bibnamefont {Baer}}, \bibinfo {author}
  {\bibfnamefont {E.}~\bibnamefont {Rabani}}, \ and\ \bibinfo {author}
  {\bibfnamefont {A.~P.}\ \bibnamefont {Alivisatos}},\ }\href {\doibase
  10.1073/pnas.1121319109} {\bibfield  {journal} {\bibinfo  {journal} {P. Natl.
  A. Sci.}\ }\textbf {\bibinfo {volume} {109}},\ \bibinfo {pages} {8016}
  (\bibinfo {year} {2012})}\BibitemShut {NoStop}%
\bibitem [{\citenamefont {Rafac}\ \emph {et~al.}(2000)\citenamefont {Rafac},
  \citenamefont {Young}, \citenamefont {Beall}, \citenamefont {Itano},
  \citenamefont {Wineland},\ and\ \citenamefont {Bergquist}}]{Rafac_2000}%
  \BibitemOpen
  \bibfield  {author} {\bibinfo {author} {\bibfnamefont {R.~J.}\ \bibnamefont
  {Rafac}}, \bibinfo {author} {\bibfnamefont {B.~C.}\ \bibnamefont {Young}},
  \bibinfo {author} {\bibfnamefont {J.~A.}\ \bibnamefont {Beall}}, \bibinfo
  {author} {\bibfnamefont {W.~M.}\ \bibnamefont {Itano}}, \bibinfo {author}
  {\bibfnamefont {D.~J.}\ \bibnamefont {Wineland}}, \ and\ \bibinfo {author}
  {\bibfnamefont {J.~C.}\ \bibnamefont {Bergquist}},\ }\href {\doibase
  10.1103/PhysRevLett.85.2462} {\bibfield  {journal} {\bibinfo  {journal}
  {Phys.\,Rev.\,Lett.}\ }\textbf {\bibinfo {volume} {85}},\ \bibinfo {pages}
  {2462} (\bibinfo {year} {2000})}\BibitemShut {NoStop}%
\bibitem [{\citenamefont {Paauw}\ \emph {et~al.}(2009)\citenamefont {Paauw},
  \citenamefont {Fedorov}, \citenamefont {Harmans},\ and\ \citenamefont
  {Mooij}}]{Paauw_2009}%
  \BibitemOpen
  \bibfield  {author} {\bibinfo {author} {\bibfnamefont {F.~G.}\ \bibnamefont
  {Paauw}}, \bibinfo {author} {\bibfnamefont {A.}~\bibnamefont {Fedorov}},
  \bibinfo {author} {\bibfnamefont {C.~J. P.~M.}\ \bibnamefont {Harmans}}, \
  and\ \bibinfo {author} {\bibfnamefont {J.~E.}\ \bibnamefont {Mooij}},\ }\href
  {\doibase 10.1103/PhysRevLett.102.090501} {\bibfield  {journal} {\bibinfo
  {journal} {Phys.\,Rev.\,Lett.}\ }\textbf {\bibinfo {volume} {102}},\ \bibinfo
  {pages} {090501} (\bibinfo {year} {2009})}\BibitemShut {NoStop}%
\bibitem [{\citenamefont {Schwarz}\ \emph {et~al.}(2013)\citenamefont
  {Schwarz}, \citenamefont {Goetz}, \citenamefont {Jiang}, \citenamefont
  {Niemczyk}, \citenamefont {Deppe}, \citenamefont {Marx},\ and\ \citenamefont
  {Gross}}]{Schwarz_2013}%
  \BibitemOpen
  \bibfield  {author} {\bibinfo {author} {\bibfnamefont {M.~J.}\ \bibnamefont
  {Schwarz}}, \bibinfo {author} {\bibfnamefont {J.}~\bibnamefont {Goetz}},
  \bibinfo {author} {\bibfnamefont {Z.}~\bibnamefont {Jiang}}, \bibinfo
  {author} {\bibfnamefont {T.}~\bibnamefont {Niemczyk}}, \bibinfo {author}
  {\bibfnamefont {F.}~\bibnamefont {Deppe}}, \bibinfo {author} {\bibfnamefont
  {A.}~\bibnamefont {Marx}}, \ and\ \bibinfo {author} {\bibfnamefont
  {R.}~\bibnamefont {Gross}},\ }\href {\doibase 10.1088/1367-2630/15/4/045001}
  {\bibfield  {journal} {\bibinfo  {journal} {New\,J.\,Phys.}\ }\textbf
  {\bibinfo {volume} {15}},\ \bibinfo {pages} {045001} (\bibinfo {year}
  {2013})}\BibitemShut {NoStop}%
\bibitem [{\citenamefont {Mostame}\ \emph {et~al.}(2012)\citenamefont
  {Mostame}, \citenamefont {Rebentrost}, \citenamefont {Eisfeld}, \citenamefont
  {Kerman}, \citenamefont {Tsomokos},\ and\ \citenamefont
  {Aspuru-Guzik}}]{Mostame_2012}%
  \BibitemOpen
  \bibfield  {author} {\bibinfo {author} {\bibfnamefont {S.}~\bibnamefont
  {Mostame}}, \bibinfo {author} {\bibfnamefont {P.}~\bibnamefont {Rebentrost}},
  \bibinfo {author} {\bibfnamefont {A.}~\bibnamefont {Eisfeld}}, \bibinfo
  {author} {\bibfnamefont {A.~J.}\ \bibnamefont {Kerman}}, \bibinfo {author}
  {\bibfnamefont {D.~I.}\ \bibnamefont {Tsomokos}}, \ and\ \bibinfo {author}
  {\bibfnamefont {A.}~\bibnamefont {Aspuru-Guzik}},\ }\href {\doibase
  10.1088/1367-2630/14/10/105013} {\bibfield  {journal} {\bibinfo  {journal}
  {New\,J.\,Phys.}\ }\textbf {\bibinfo {volume} {14}},\ \bibinfo {pages}
  {105013} (\bibinfo {year} {2012})}\BibitemShut {NoStop}%
\bibitem [{\citenamefont {Mostame}\ \emph {et~al.}(2016)\citenamefont
  {Mostame}, \citenamefont {Huh}, \citenamefont {Kreisbeck}, \citenamefont
  {Kerman}, \citenamefont {Fujita}, \citenamefont {Eisfeld},\ and\
  \citenamefont {Aspuru-Guzik}}]{Mostame_2016}%
  \BibitemOpen
  \bibfield  {author} {\bibinfo {author} {\bibfnamefont {S.}~\bibnamefont
  {Mostame}}, \bibinfo {author} {\bibfnamefont {J.}~\bibnamefont {Huh}},
  \bibinfo {author} {\bibfnamefont {C.}~\bibnamefont {Kreisbeck}}, \bibinfo
  {author} {\bibfnamefont {A.~J.}\ \bibnamefont {Kerman}}, \bibinfo {author}
  {\bibfnamefont {T.}~\bibnamefont {Fujita}}, \bibinfo {author} {\bibfnamefont
  {A.}~\bibnamefont {Eisfeld}}, \ and\ \bibinfo {author} {\bibfnamefont
  {A.}~\bibnamefont {Aspuru-Guzik}},\ }\href {\doibase
  10.1007/s11128-016-1489-3} {\bibfield  {journal} {\bibinfo  {journal}
  {Quantum Inf. Process.}\ }\textbf {\bibinfo {volume} {16}},\ \bibinfo {pages}
  {44} (\bibinfo {year} {2016})}\BibitemShut {NoStop}%
\bibitem [{\citenamefont {Srinivasan}\ \emph {et~al.}(2011)\citenamefont
  {Srinivasan}, \citenamefont {Hoffman}, \citenamefont {Gambetta},\ and\
  \citenamefont {Houck}}]{Srinivasan_2011}%
  \BibitemOpen
  \bibfield  {author} {\bibinfo {author} {\bibfnamefont {S.~J.}\ \bibnamefont
  {Srinivasan}}, \bibinfo {author} {\bibfnamefont {A.~J.}\ \bibnamefont
  {Hoffman}}, \bibinfo {author} {\bibfnamefont {J.~M.}\ \bibnamefont
  {Gambetta}}, \ and\ \bibinfo {author} {\bibfnamefont {A.~A.}\ \bibnamefont
  {Houck}},\ }\href {\doibase 10.1103/PhysRevLett.106.083601} {\bibfield
  {journal} {\bibinfo  {journal} {Phys.\,Rev.\,Lett.}\ }\textbf {\bibinfo
  {volume} {106}},\ \bibinfo {pages} {083601} (\bibinfo {year}
  {2011})}\BibitemShut {NoStop}%
\bibitem [{\citenamefont {You}\ and\ \citenamefont {Nori}(2011)}]{You_2011}%
  \BibitemOpen
  \bibfield  {author} {\bibinfo {author} {\bibfnamefont {J.~Q.}\ \bibnamefont
  {You}}\ and\ \bibinfo {author} {\bibfnamefont {F.}~\bibnamefont {Nori}},\
  }\href {\doibase 10.1038/nature10122} {\bibfield  {journal} {\bibinfo
  {journal} {Nature}\ }\textbf {\bibinfo {volume} {474}},\ \bibinfo {pages}
  {589} (\bibinfo {year} {2011})}\BibitemShut {NoStop}%
\bibitem [{\citenamefont {de~Groot}\ \emph {et~al.}(2010)\citenamefont
  {de~Groot}, \citenamefont {Lisenfeld}, \citenamefont {Schouten},
  \citenamefont {Ashhab}, \citenamefont {Lupascu}, \citenamefont {Harmans},\
  and\ \citenamefont {Mooij}}]{deGroot_2010}%
  \BibitemOpen
  \bibfield  {author} {\bibinfo {author} {\bibfnamefont {P.~C.}\ \bibnamefont
  {de~Groot}}, \bibinfo {author} {\bibfnamefont {J.}~\bibnamefont {Lisenfeld}},
  \bibinfo {author} {\bibfnamefont {R.~N.}\ \bibnamefont {Schouten}}, \bibinfo
  {author} {\bibfnamefont {S.}~\bibnamefont {Ashhab}}, \bibinfo {author}
  {\bibfnamefont {A.}~\bibnamefont {Lupascu}}, \bibinfo {author} {\bibfnamefont
  {C.~J. P.~M.}\ \bibnamefont {Harmans}}, \ and\ \bibinfo {author}
  {\bibfnamefont {J.~E.}\ \bibnamefont {Mooij}},\ }\href {\doibase
  10.1038/nphys1733} {\bibfield  {journal} {\bibinfo  {journal} {Nat.\,Phys.}\
  }\textbf {\bibinfo {volume} {6}},\ \bibinfo {pages} {763} (\bibinfo {year}
  {2010})}\BibitemShut {NoStop}%
\bibitem [{\citenamefont {Didier}\ \emph {et~al.}(2015)\citenamefont {Didier},
  \citenamefont {Bourassa},\ and\ \citenamefont {Blais}}]{Didier_2015}%
  \BibitemOpen
  \bibfield  {author} {\bibinfo {author} {\bibfnamefont {N.}~\bibnamefont
  {Didier}}, \bibinfo {author} {\bibfnamefont {J.}~\bibnamefont {Bourassa}}, \
  and\ \bibinfo {author} {\bibfnamefont {A.}~\bibnamefont {Blais}},\ }\href
  {\doibase 10.1103/PhysRevLett.115.203601} {\bibfield  {journal} {\bibinfo
  {journal} {Phys.\,Rev.\,Lett.}\ }\textbf {\bibinfo {volume} {115}},\ \bibinfo
  {pages} {203601} (\bibinfo {year} {2015})}\BibitemShut {NoStop}%
\bibitem [{\citenamefont {Forn-D{\'i}az}\ \emph {et~al.}(2016)\citenamefont
  {Forn-D{\'i}az}, \citenamefont {Romero}, \citenamefont {Harmans},
  \citenamefont {Solano},\ and\ \citenamefont {Mooij}}]{Forn-Diaz_2015}%
  \BibitemOpen
  \bibfield  {author} {\bibinfo {author} {\bibfnamefont {P.}~\bibnamefont
  {Forn-D{\'i}az}}, \bibinfo {author} {\bibfnamefont {G.}~\bibnamefont
  {Romero}}, \bibinfo {author} {\bibfnamefont {C.~J. P.~M.}\ \bibnamefont
  {Harmans}}, \bibinfo {author} {\bibfnamefont {E.}~\bibnamefont {Solano}}, \
  and\ \bibinfo {author} {\bibfnamefont {J.~E.}\ \bibnamefont {Mooij}},\ }\href
  {\doibase 10.1038/srep26720} {\bibfield  {journal} {\bibinfo  {journal}
  {Sci.\,Rep.}\ }\textbf {\bibinfo {volume} {6}},\ \bibinfo {pages} {26720}
  (\bibinfo {year} {2016})}\BibitemShut {NoStop}%
\bibitem [{\citenamefont {{Wu}}\ \emph {et~al.}(2016)\citenamefont {{Wu}},
  \citenamefont {{Yang}}, \citenamefont {{Zheng}}, \citenamefont {{Deng}},
  \citenamefont {{Yan}}, \citenamefont {{Zhao}}, \citenamefont {{Huang}},
  \citenamefont {{Munro}}, \citenamefont {{Nemoto}}, \citenamefont {{Zheng}},
  \citenamefont {{Sun}}, \citenamefont {{Liu}}, \citenamefont {{Zhu}},\ and\
  \citenamefont {{Lu}}}]{Wu_2016}%
  \BibitemOpen
  \bibfield  {author} {\bibinfo {author} {\bibfnamefont {Y.}~\bibnamefont
  {{Wu}}}, \bibinfo {author} {\bibfnamefont {L.-P.}\ \bibnamefont {{Yang}}},
  \bibinfo {author} {\bibfnamefont {Y.}~\bibnamefont {{Zheng}}}, \bibinfo
  {author} {\bibfnamefont {H.}~\bibnamefont {{Deng}}}, \bibinfo {author}
  {\bibfnamefont {Z.}~\bibnamefont {{Yan}}}, \bibinfo {author} {\bibfnamefont
  {Y.}~\bibnamefont {{Zhao}}}, \bibinfo {author} {\bibfnamefont
  {K.}~\bibnamefont {{Huang}}}, \bibinfo {author} {\bibfnamefont {W.~J.}\
  \bibnamefont {{Munro}}}, \bibinfo {author} {\bibfnamefont {K.}~\bibnamefont
  {{Nemoto}}}, \bibinfo {author} {\bibfnamefont {D.-N.}\ \bibnamefont
  {{Zheng}}}, \bibinfo {author} {\bibfnamefont {C.~P.}\ \bibnamefont {{Sun}}},
  \bibinfo {author} {\bibfnamefont {Y.-x.}\ \bibnamefont {{Liu}}}, \bibinfo
  {author} {\bibfnamefont {X.}~\bibnamefont {{Zhu}}}, \ and\ \bibinfo {author}
  {\bibfnamefont {L.}~\bibnamefont {{Lu}}},\ }\href@noop {} {\bibfield
  {journal} {\bibinfo  {journal} {ArXiv e-prints}\ } (\bibinfo {year}
  {2016})},\ \Eprint {http://arxiv.org/abs/1605.06747} {arXiv:1605.06747
  [quant-ph]} \BibitemShut {NoStop}%
\bibitem [{\citenamefont {Royer}\ \emph {et~al.}(2017)\citenamefont {Royer},
  \citenamefont {Grimsmo}, \citenamefont {Didier},\ and\ \citenamefont
  {Blais}}]{Royer_2017}%
  \BibitemOpen
  \bibfield  {author} {\bibinfo {author} {\bibfnamefont {B.}~\bibnamefont
  {Royer}}, \bibinfo {author} {\bibfnamefont {A.~L.}\ \bibnamefont {Grimsmo}},
  \bibinfo {author} {\bibfnamefont {N.}~\bibnamefont {Didier}}, \ and\ \bibinfo
  {author} {\bibfnamefont {A.}~\bibnamefont {Blais}},\ }\href {\doibase
  10.22331/q-2017-05-11-11} {\bibfield  {journal} {\bibinfo  {journal}
  {{Quantum}}\ }\textbf {\bibinfo {volume} {1}},\ \bibinfo {pages} {11}
  (\bibinfo {year} {2017})}\BibitemShut {NoStop}%
\bibitem [{\citenamefont {Vool}\ \emph {et~al.}(2018)\citenamefont {Vool},
  \citenamefont {Kou}, \citenamefont {Smith}, \citenamefont {Frattini},
  \citenamefont {Serniak}, \citenamefont {Reinhold}, \citenamefont {Pop},
  \citenamefont {Shankar}, \citenamefont {Frunzio}, \citenamefont {Girvin},\
  and\ \citenamefont {Devoret}}]{Vool_2018}%
  \BibitemOpen
  \bibfield  {author} {\bibinfo {author} {\bibfnamefont {U.}~\bibnamefont
  {Vool}}, \bibinfo {author} {\bibfnamefont {A.}~\bibnamefont {Kou}}, \bibinfo
  {author} {\bibfnamefont {W.~C.}\ \bibnamefont {Smith}}, \bibinfo {author}
  {\bibfnamefont {N.~E.}\ \bibnamefont {Frattini}}, \bibinfo {author}
  {\bibfnamefont {K.}~\bibnamefont {Serniak}}, \bibinfo {author} {\bibfnamefont
  {P.}~\bibnamefont {Reinhold}}, \bibinfo {author} {\bibfnamefont {I.~M.}\
  \bibnamefont {Pop}}, \bibinfo {author} {\bibfnamefont {S.}~\bibnamefont
  {Shankar}}, \bibinfo {author} {\bibfnamefont {L.}~\bibnamefont {Frunzio}},
  \bibinfo {author} {\bibfnamefont {S.~M.}\ \bibnamefont {Girvin}}, \ and\
  \bibinfo {author} {\bibfnamefont {M.~H.}\ \bibnamefont {Devoret}},\ }\href
  {\doibase 10.1103/PhysRevApplied.9.054046} {\bibfield  {journal} {\bibinfo
  {journal} {Phys. Rev. Applied}\ }\textbf {\bibinfo {volume} {9}},\ \bibinfo
  {pages} {054046} (\bibinfo {year} {2018})}\BibitemShut {NoStop}%
\bibitem [{\citenamefont {Billangeon}\ \emph {et~al.}(2015)\citenamefont
  {Billangeon}, \citenamefont {Tsai},\ and\ \citenamefont
  {Nakamura}}]{Billangeon_2015}%
  \BibitemOpen
  \bibfield  {author} {\bibinfo {author} {\bibfnamefont {P.-M.}\ \bibnamefont
  {Billangeon}}, \bibinfo {author} {\bibfnamefont {J.~S.}\ \bibnamefont
  {Tsai}}, \ and\ \bibinfo {author} {\bibfnamefont {Y.}~\bibnamefont
  {Nakamura}},\ }\href {\doibase 10.1103/PhysRevB.91.094517} {\bibfield
  {journal} {\bibinfo  {journal} {Phys.\,Rev.\,B}\ }\textbf {\bibinfo {volume}
  {91}},\ \bibinfo {pages} {094517} (\bibinfo {year} {2015})}\BibitemShut
  {NoStop}%
\bibitem [{\citenamefont {Liu}\ \emph {et~al.}(2014)\citenamefont {Liu},
  \citenamefont {Yang}, \citenamefont {Sun},\ and\ \citenamefont
  {Wang}}]{Liu_2014}%
  \BibitemOpen
  \bibfield  {author} {\bibinfo {author} {\bibfnamefont {Y.}~\bibnamefont
  {Liu}}, \bibinfo {author} {\bibfnamefont {C.-X.}\ \bibnamefont {Yang}},
  \bibinfo {author} {\bibfnamefont {H.-C.}\ \bibnamefont {Sun}}, \ and\
  \bibinfo {author} {\bibfnamefont {X.-B.}\ \bibnamefont {Wang}},\ }\href
  {\doibase 10.1088/1367-2630/16/1/015031} {\bibfield  {journal} {\bibinfo
  {journal} {New\,J.\,Phys.}\ }\textbf {\bibinfo {volume} {16}},\ \bibinfo
  {pages} {015031} (\bibinfo {year} {2014})}\BibitemShut {NoStop}%
\bibitem [{\citenamefont {Blais}\ \emph {et~al.}(2007)\citenamefont {Blais},
  \citenamefont {Gambetta}, \citenamefont {Wallraff}, \citenamefont {Schuster},
  \citenamefont {Girvin}, \citenamefont {Devoret},\ and\ \citenamefont
  {Schoelkopf}}]{Blais_2007}%
  \BibitemOpen
  \bibfield  {author} {\bibinfo {author} {\bibfnamefont {A.}~\bibnamefont
  {Blais}}, \bibinfo {author} {\bibfnamefont {J.}~\bibnamefont {Gambetta}},
  \bibinfo {author} {\bibfnamefont {A.}~\bibnamefont {Wallraff}}, \bibinfo
  {author} {\bibfnamefont {D.~I.}\ \bibnamefont {Schuster}}, \bibinfo {author}
  {\bibfnamefont {S.~M.}\ \bibnamefont {Girvin}}, \bibinfo {author}
  {\bibfnamefont {M.~H.}\ \bibnamefont {Devoret}}, \ and\ \bibinfo {author}
  {\bibfnamefont {R.~J.}\ \bibnamefont {Schoelkopf}},\ }\href {\doibase
  10.1103/PhysRevA.75.032329} {\bibfield  {journal} {\bibinfo  {journal}
  {Phys.\,Rev.\,A}\ }\textbf {\bibinfo {volume} {75}},\ \bibinfo {pages}
  {032329} (\bibinfo {year} {2007})}\BibitemShut {NoStop}%
\bibitem [{sup()}]{supp}%
  \BibitemOpen
  \href@noop {} {}\bibinfo {note} {See Supplemental Material for experimental
  techniques and theoretical methods, which includes
  Refs.~\onlinecite{Abragam_1961,Orlando_1999,Schuster_2005,Raab_2005,Cohen_2006,Gambetta_2008,Boissonneault_2009,Niemczyk_2010,Niemczyk_2011,Kasperczyk_2015,Goetz_2016b}}\BibitemShut
  {NoStop}%
\bibitem [{\citenamefont {Goetz}\ \emph {et~al.}(2016)\citenamefont {Goetz},
  \citenamefont {Deppe}, \citenamefont {Haeberlein}, \citenamefont {Wulschner},
  \citenamefont {Zollitsch}, \citenamefont {Meier}, \citenamefont {Fischer},
  \citenamefont {Eder}, \citenamefont {Xie}, \citenamefont {Fedorov},
  \citenamefont {Menzel}, \citenamefont {Marx},\ and\ \citenamefont
  {Gross}}]{Goetz_2016}%
  \BibitemOpen
  \bibfield  {author} {\bibinfo {author} {\bibfnamefont {J.}~\bibnamefont
  {Goetz}}, \bibinfo {author} {\bibfnamefont {F.}~\bibnamefont {Deppe}},
  \bibinfo {author} {\bibfnamefont {M.}~\bibnamefont {Haeberlein}}, \bibinfo
  {author} {\bibfnamefont {F.}~\bibnamefont {Wulschner}}, \bibinfo {author}
  {\bibfnamefont {C.~W.}\ \bibnamefont {Zollitsch}}, \bibinfo {author}
  {\bibfnamefont {S.}~\bibnamefont {Meier}}, \bibinfo {author} {\bibfnamefont
  {M.}~\bibnamefont {Fischer}}, \bibinfo {author} {\bibfnamefont
  {P.}~\bibnamefont {Eder}}, \bibinfo {author} {\bibfnamefont {E.}~\bibnamefont
  {Xie}}, \bibinfo {author} {\bibfnamefont {K.~G.}\ \bibnamefont {Fedorov}},
  \bibinfo {author} {\bibfnamefont {E.~P.}\ \bibnamefont {Menzel}}, \bibinfo
  {author} {\bibfnamefont {A.}~\bibnamefont {Marx}}, \ and\ \bibinfo {author}
  {\bibfnamefont {R.}~\bibnamefont {Gross}},\ }\href {\doibase
  10.1063/1.4939299} {\bibfield  {journal} {\bibinfo  {journal}
  {J.\,App.\,Phys.}\ }\textbf {\bibinfo {volume} {119}},\ \bibinfo {pages}
  {015304} (\bibinfo {year} {2016})}\BibitemShut {NoStop}%
\bibitem [{\citenamefont {Dolan}(1977)}]{Dolan_1977}%
  \BibitemOpen
  \bibfield  {author} {\bibinfo {author} {\bibfnamefont {G.~J.}\ \bibnamefont
  {Dolan}},\ }\href {\doibase 10.1063/1.89690} {\bibfield  {journal} {\bibinfo
  {journal} {Appl.\,Phys.\,Lett.}\ }\textbf {\bibinfo {volume} {31}},\ \bibinfo
  {pages} {337} (\bibinfo {year} {1977})}\BibitemShut {NoStop}%
\bibitem [{\citenamefont {Goetz}\ \emph
  {et~al.}(2017{\natexlab{a}})\citenamefont {Goetz}, \citenamefont
  {Pogorzalek}, \citenamefont {Deppe}, \citenamefont {Fedorov}, \citenamefont
  {Eder}, \citenamefont {Fischer}, \citenamefont {Wulschner}, \citenamefont
  {Xie}, \citenamefont {Marx},\ and\ \citenamefont {Gross}}]{Goetz_2016a}%
  \BibitemOpen
  \bibfield  {author} {\bibinfo {author} {\bibfnamefont {J.}~\bibnamefont
  {Goetz}}, \bibinfo {author} {\bibfnamefont {S.}~\bibnamefont {Pogorzalek}},
  \bibinfo {author} {\bibfnamefont {F.}~\bibnamefont {Deppe}}, \bibinfo
  {author} {\bibfnamefont {K.~G.}\ \bibnamefont {Fedorov}}, \bibinfo {author}
  {\bibfnamefont {P.}~\bibnamefont {Eder}}, \bibinfo {author} {\bibfnamefont
  {M.}~\bibnamefont {Fischer}}, \bibinfo {author} {\bibfnamefont
  {F.}~\bibnamefont {Wulschner}}, \bibinfo {author} {\bibfnamefont
  {E.}~\bibnamefont {Xie}}, \bibinfo {author} {\bibfnamefont {A.}~\bibnamefont
  {Marx}}, \ and\ \bibinfo {author} {\bibfnamefont {R.}~\bibnamefont {Gross}},\
  }\href {\doibase 10.1103/PhysRevLett.118.103602} {\bibfield  {journal}
  {\bibinfo  {journal} {Phys.\,Rev.\,Lett.}\ }\textbf {\bibinfo {volume}
  {118}},\ \bibinfo {pages} {103602} (\bibinfo {year}
  {2017}{\natexlab{a}})}\BibitemShut {NoStop}%
\bibitem [{\citenamefont {Yan}\ \emph {et~al.}(2016)\citenamefont {Yan},
  \citenamefont {Gustavsson}, \citenamefont {Kamal}, \citenamefont {Birenbaum},
  \citenamefont {Sears}, \citenamefont {Hover}, \citenamefont {Gudmundsen},
  \citenamefont {Rosenberg}, \citenamefont {Samach}, \citenamefont {Weber},
  \citenamefont {Yoder}, \citenamefont {Orlando}, \citenamefont {Clarke},
  \citenamefont {Kerman},\ and\ \citenamefont {Oliver}}]{Yan_2015}%
  \BibitemOpen
  \bibfield  {author} {\bibinfo {author} {\bibfnamefont {F.}~\bibnamefont
  {Yan}}, \bibinfo {author} {\bibfnamefont {S.}~\bibnamefont {Gustavsson}},
  \bibinfo {author} {\bibfnamefont {A.}~\bibnamefont {Kamal}}, \bibinfo
  {author} {\bibfnamefont {J.}~\bibnamefont {Birenbaum}}, \bibinfo {author}
  {\bibfnamefont {A.~P.}\ \bibnamefont {Sears}}, \bibinfo {author}
  {\bibfnamefont {D.}~\bibnamefont {Hover}}, \bibinfo {author} {\bibfnamefont
  {T.~J.}\ \bibnamefont {Gudmundsen}}, \bibinfo {author} {\bibfnamefont
  {D.}~\bibnamefont {Rosenberg}}, \bibinfo {author} {\bibfnamefont
  {G.}~\bibnamefont {Samach}}, \bibinfo {author} {\bibfnamefont
  {S.}~\bibnamefont {Weber}}, \bibinfo {author} {\bibfnamefont {J.~L.}\
  \bibnamefont {Yoder}}, \bibinfo {author} {\bibfnamefont {T.~P.}\ \bibnamefont
  {Orlando}}, \bibinfo {author} {\bibfnamefont {J.}~\bibnamefont {Clarke}},
  \bibinfo {author} {\bibfnamefont {A.~J.}\ \bibnamefont {Kerman}}, \ and\
  \bibinfo {author} {\bibfnamefont {W.~D.}\ \bibnamefont {Oliver}},\ }\href
  {\doibase 10.1038/ncomms12964} {\bibfield  {journal} {\bibinfo  {journal}
  {Nat.\,Comm.}\ }\textbf {\bibinfo {volume} {7}},\ \bibinfo {pages} {12964}
  (\bibinfo {year} {2016})}\BibitemShut {NoStop}%
\bibitem [{\citenamefont {Tan}\ \emph {et~al.}(2017)\citenamefont {Tan},
  \citenamefont {Partanen}, \citenamefont {Lake}, \citenamefont {Govenius},
  \citenamefont {Masuda},\ and\ \citenamefont {M{\"o}tt{\"o}nen}}]{Tan_2017}%
  \BibitemOpen
  \bibfield  {author} {\bibinfo {author} {\bibfnamefont {K.~Y.}\ \bibnamefont
  {Tan}}, \bibinfo {author} {\bibfnamefont {M.}~\bibnamefont {Partanen}},
  \bibinfo {author} {\bibfnamefont {R.~E.}\ \bibnamefont {Lake}}, \bibinfo
  {author} {\bibfnamefont {J.}~\bibnamefont {Govenius}}, \bibinfo {author}
  {\bibfnamefont {S.}~\bibnamefont {Masuda}}, \ and\ \bibinfo {author}
  {\bibfnamefont {M.}~\bibnamefont {M{\"o}tt{\"o}nen}},\ }\href {\doibase
  10.1038/ncomms15189} {\bibfield  {journal} {\bibinfo  {journal}
  {Nat.\,Comm.}\ }\textbf {\bibinfo {volume} {8}},\ \bibinfo {pages} {15189}
  (\bibinfo {year} {2017})}\BibitemShut {NoStop}%
\bibitem [{\citenamefont {{Poto{\v c}nik}}\ \emph {et~al.}(2018)\citenamefont
  {{Poto{\v c}nik}}, \citenamefont {{Bargerbos}}, \citenamefont
  {{Schr{\"o}der}}, \citenamefont {{Khan}}, \citenamefont {{Collodo}},
  \citenamefont {{Gasparinetti}}, \citenamefont {{Salath{\'e}}}, \citenamefont
  {{Creatore}}, \citenamefont {{Eichler}}, \citenamefont {{T{\"u}reci}},
  \citenamefont {{Chin}},\ and\ \citenamefont {{Wallraff}}}]{Potocnik_2018}%
  \BibitemOpen
  \bibfield  {author} {\bibinfo {author} {\bibfnamefont {A.}~\bibnamefont
  {{Poto{\v c}nik}}}, \bibinfo {author} {\bibfnamefont {A.}~\bibnamefont
  {{Bargerbos}}}, \bibinfo {author} {\bibfnamefont {F.~A.~Y.~N.}\ \bibnamefont
  {{Schr{\"o}der}}}, \bibinfo {author} {\bibfnamefont {S.~A.}\ \bibnamefont
  {{Khan}}}, \bibinfo {author} {\bibfnamefont {M.~C.}\ \bibnamefont
  {{Collodo}}}, \bibinfo {author} {\bibfnamefont {S.}~\bibnamefont
  {{Gasparinetti}}}, \bibinfo {author} {\bibfnamefont {Y.}~\bibnamefont
  {{Salath{\'e}}}}, \bibinfo {author} {\bibfnamefont {C.}~\bibnamefont
  {{Creatore}}}, \bibinfo {author} {\bibfnamefont {C.}~\bibnamefont
  {{Eichler}}}, \bibinfo {author} {\bibfnamefont {H.~E.}\ \bibnamefont
  {{T{\"u}reci}}}, \bibinfo {author} {\bibfnamefont {A.~W.}\ \bibnamefont
  {{Chin}}}, \ and\ \bibinfo {author} {\bibfnamefont {A.}~\bibnamefont
  {{Wallraff}}},\ }\href {\doibase doi.org/10.1038/s41467-018-03312-x}
  {\bibfield  {journal} {\bibinfo  {journal} {Nat.\,Comm.}\ }\textbf {\bibinfo
  {volume} {9}},\ \bibinfo {pages} {904} (\bibinfo {year} {2018})}\BibitemShut
  {NoStop}%
\bibitem [{\citenamefont {Hormozi}\ \emph {et~al.}(2017)\citenamefont
  {Hormozi}, \citenamefont {Brown}, \citenamefont {Carleo},\ and\ \citenamefont
  {Troyer}}]{Hormozi_2017}%
  \BibitemOpen
  \bibfield  {author} {\bibinfo {author} {\bibfnamefont {L.}~\bibnamefont
  {Hormozi}}, \bibinfo {author} {\bibfnamefont {E.~W.}\ \bibnamefont {Brown}},
  \bibinfo {author} {\bibfnamefont {G.}~\bibnamefont {Carleo}}, \ and\ \bibinfo
  {author} {\bibfnamefont {M.}~\bibnamefont {Troyer}},\ }\href {\doibase
  10.1103/PhysRevB.95.184416} {\bibfield  {journal} {\bibinfo  {journal} {Phys.
  Rev. B}\ }\textbf {\bibinfo {volume} {95}},\ \bibinfo {pages} {184416}
  (\bibinfo {year} {2017})}\BibitemShut {NoStop}%
\bibitem [{\citenamefont {Albash}\ and\ \citenamefont
  {Lidar}(2018)}]{Albash_2018}%
  \BibitemOpen
  \bibfield  {author} {\bibinfo {author} {\bibfnamefont {T.}~\bibnamefont
  {Albash}}\ and\ \bibinfo {author} {\bibfnamefont {D.~A.}\ \bibnamefont
  {Lidar}},\ }\href {\doibase 10.1103/RevModPhys.90.015002} {\bibfield
  {journal} {\bibinfo  {journal} {Rev.\,Mod.\,Phys.}\ }\textbf {\bibinfo
  {volume} {90}},\ \bibinfo {pages} {015002} (\bibinfo {year}
  {2018})}\BibitemShut {NoStop}%
\bibitem [{\citenamefont {Brooke}\ \emph {et~al.}(1999)\citenamefont {Brooke},
  \citenamefont {Bitko}, \citenamefont {Rosenbaum},\ and\ \citenamefont
  {Aeppli}}]{Brooke_1999}%
  \BibitemOpen
  \bibfield  {author} {\bibinfo {author} {\bibfnamefont {J.}~\bibnamefont
  {Brooke}}, \bibinfo {author} {\bibfnamefont {D.}~\bibnamefont {Bitko}},
  \bibinfo {author} {\bibfnamefont {F.~T.}\ \bibnamefont {Rosenbaum}}, \ and\
  \bibinfo {author} {\bibfnamefont {G.}~\bibnamefont {Aeppli}},\ }\href
  {\doibase 10.1126/science.284.5415.779} {\bibfield  {journal} {\bibinfo
  {journal} {Science}\ }\textbf {\bibinfo {volume} {284}},\ \bibinfo {pages}
  {779} (\bibinfo {year} {1999})}\BibitemShut {NoStop}%
\bibitem [{\citenamefont {Aharonov}\ \emph {et~al.}(2007)\citenamefont
  {Aharonov}, \citenamefont {van Dam}, \citenamefont {Kempe}, \citenamefont
  {Landau}, \citenamefont {Lloyd},\ and\ \citenamefont {Regev}}]{Dorit_2007}%
  \BibitemOpen
  \bibfield  {author} {\bibinfo {author} {\bibfnamefont {D.}~\bibnamefont
  {Aharonov}}, \bibinfo {author} {\bibfnamefont {W.}~\bibnamefont {van Dam}},
  \bibinfo {author} {\bibfnamefont {J.}~\bibnamefont {Kempe}}, \bibinfo
  {author} {\bibfnamefont {Z.}~\bibnamefont {Landau}}, \bibinfo {author}
  {\bibfnamefont {S.}~\bibnamefont {Lloyd}}, \ and\ \bibinfo {author}
  {\bibfnamefont {O.}~\bibnamefont {Regev}},\ }\href {\doibase
  10.1137/S0097539705447323} {\bibfield  {journal} {\bibinfo  {journal} {SIAM
  Journal on Computing}\ }\textbf {\bibinfo {volume} {37}},\ \bibinfo {pages}
  {166} (\bibinfo {year} {2007})}\BibitemShut {NoStop}%
\bibitem [{\citenamefont {Lanting}\ \emph {et~al.}(2014)\citenamefont
  {Lanting}, \citenamefont {Przybysz}, \citenamefont {Smirnov}, \citenamefont
  {Spedalieri}, \citenamefont {Amin}, \citenamefont {Berkley}, \citenamefont
  {Harris}, \citenamefont {Altomare}, \citenamefont {Boixo}, \citenamefont
  {Bunyk}, \citenamefont {Dickson}, \citenamefont {Enderud}, \citenamefont
  {Hilton}, \citenamefont {Hoskinson}, \citenamefont {Johnson}, \citenamefont
  {Ladizinsky}, \citenamefont {Ladizinsky}, \citenamefont {Neufeld},
  \citenamefont {Oh}, \citenamefont {Perminov}, \citenamefont {Rich},
  \citenamefont {Thom}, \citenamefont {Tolkacheva}, \citenamefont {Uchaikin},
  \citenamefont {Wilson},\ and\ \citenamefont {Rose}}]{Lanting_2014}%
  \BibitemOpen
  \bibfield  {author} {\bibinfo {author} {\bibfnamefont {T.}~\bibnamefont
  {Lanting}}, \bibinfo {author} {\bibfnamefont {A.~J.}\ \bibnamefont
  {Przybysz}}, \bibinfo {author} {\bibfnamefont {A.~Y.}\ \bibnamefont
  {Smirnov}}, \bibinfo {author} {\bibfnamefont {F.~M.}\ \bibnamefont
  {Spedalieri}}, \bibinfo {author} {\bibfnamefont {M.~H.}\ \bibnamefont
  {Amin}}, \bibinfo {author} {\bibfnamefont {A.~J.}\ \bibnamefont {Berkley}},
  \bibinfo {author} {\bibfnamefont {R.}~\bibnamefont {Harris}}, \bibinfo
  {author} {\bibfnamefont {F.}~\bibnamefont {Altomare}}, \bibinfo {author}
  {\bibfnamefont {S.}~\bibnamefont {Boixo}}, \bibinfo {author} {\bibfnamefont
  {P.}~\bibnamefont {Bunyk}}, \bibinfo {author} {\bibfnamefont
  {N.}~\bibnamefont {Dickson}}, \bibinfo {author} {\bibfnamefont
  {C.}~\bibnamefont {Enderud}}, \bibinfo {author} {\bibfnamefont {J.~P.}\
  \bibnamefont {Hilton}}, \bibinfo {author} {\bibfnamefont {E.}~\bibnamefont
  {Hoskinson}}, \bibinfo {author} {\bibfnamefont {M.~W.}\ \bibnamefont
  {Johnson}}, \bibinfo {author} {\bibfnamefont {E.}~\bibnamefont {Ladizinsky}},
  \bibinfo {author} {\bibfnamefont {N.}~\bibnamefont {Ladizinsky}}, \bibinfo
  {author} {\bibfnamefont {R.}~\bibnamefont {Neufeld}}, \bibinfo {author}
  {\bibfnamefont {T.}~\bibnamefont {Oh}}, \bibinfo {author} {\bibfnamefont
  {I.}~\bibnamefont {Perminov}}, \bibinfo {author} {\bibfnamefont
  {C.}~\bibnamefont {Rich}}, \bibinfo {author} {\bibfnamefont {M.~C.}\
  \bibnamefont {Thom}}, \bibinfo {author} {\bibfnamefont {E.}~\bibnamefont
  {Tolkacheva}}, \bibinfo {author} {\bibfnamefont {S.}~\bibnamefont
  {Uchaikin}}, \bibinfo {author} {\bibfnamefont {A.~B.}\ \bibnamefont
  {Wilson}}, \ and\ \bibinfo {author} {\bibfnamefont {G.}~\bibnamefont
  {Rose}},\ }\href {\doibase 10.1103/PhysRevX.4.021041} {\bibfield  {journal}
  {\bibinfo  {journal} {Phys.\,Rev.\,X}\ }\textbf {\bibinfo {volume} {4}},\
  \bibinfo {pages} {021041} (\bibinfo {year} {2014})}\BibitemShut {NoStop}%
\bibitem [{\citenamefont {Vinci}\ and\ \citenamefont
  {Lidar}(2017)}]{Vinci_2017}%
  \BibitemOpen
  \bibfield  {author} {\bibinfo {author} {\bibfnamefont {W.}~\bibnamefont
  {Vinci}}\ and\ \bibinfo {author} {\bibfnamefont {D.~A.}\ \bibnamefont
  {Lidar}},\ }\href {\doibase 10.1038/s41534-017-0037-z} {\bibfield  {journal}
  {\bibinfo  {journal} {npj Quantum Information}\ }\textbf {\bibinfo {volume}
  {3}},\ \bibinfo {pages} {38} (\bibinfo {year} {2017})}\BibitemShut {NoStop}%
\bibitem [{\citenamefont {Baust}\ \emph {et~al.}(2015)\citenamefont {Baust},
  \citenamefont {Hoffmann}, \citenamefont {Haeberlein}, \citenamefont
  {Schwarz}, \citenamefont {Eder}, \citenamefont {Goetz}, \citenamefont
  {Wulschner}, \citenamefont {Xie}, \citenamefont {Zhong}, \citenamefont
  {Quijandr\'{\i}a}, \citenamefont {Peropadre}, \citenamefont {Zueco},
  \citenamefont {Garc\'{\i}a~Ripoll}, \citenamefont {Solano}, \citenamefont
  {Fedorov}, \citenamefont {Menzel}, \citenamefont {Deppe}, \citenamefont
  {Marx},\ and\ \citenamefont {Gross}}]{Baust_2015}%
  \BibitemOpen
  \bibfield  {author} {\bibinfo {author} {\bibfnamefont {A.}~\bibnamefont
  {Baust}}, \bibinfo {author} {\bibfnamefont {E.}~\bibnamefont {Hoffmann}},
  \bibinfo {author} {\bibfnamefont {M.}~\bibnamefont {Haeberlein}}, \bibinfo
  {author} {\bibfnamefont {M.~J.}\ \bibnamefont {Schwarz}}, \bibinfo {author}
  {\bibfnamefont {P.}~\bibnamefont {Eder}}, \bibinfo {author} {\bibfnamefont
  {J.}~\bibnamefont {Goetz}}, \bibinfo {author} {\bibfnamefont
  {F.}~\bibnamefont {Wulschner}}, \bibinfo {author} {\bibfnamefont
  {E.}~\bibnamefont {Xie}}, \bibinfo {author} {\bibfnamefont {L.}~\bibnamefont
  {Zhong}}, \bibinfo {author} {\bibfnamefont {F.}~\bibnamefont
  {Quijandr\'{\i}a}}, \bibinfo {author} {\bibfnamefont {B.}~\bibnamefont
  {Peropadre}}, \bibinfo {author} {\bibfnamefont {D.}~\bibnamefont {Zueco}},
  \bibinfo {author} {\bibfnamefont {J.-J.}\ \bibnamefont {Garc\'{\i}a~Ripoll}},
  \bibinfo {author} {\bibfnamefont {E.}~\bibnamefont {Solano}}, \bibinfo
  {author} {\bibfnamefont {K.}~\bibnamefont {Fedorov}}, \bibinfo {author}
  {\bibfnamefont {E.~P.}\ \bibnamefont {Menzel}}, \bibinfo {author}
  {\bibfnamefont {F.}~\bibnamefont {Deppe}}, \bibinfo {author} {\bibfnamefont
  {A.}~\bibnamefont {Marx}}, \ and\ \bibinfo {author} {\bibfnamefont
  {R.}~\bibnamefont {Gross}},\ }\href {\doibase 10.1103/PhysRevB.91.014515}
  {\bibfield  {journal} {\bibinfo  {journal} {Phys.\,Rev.\,B}\ }\textbf
  {\bibinfo {volume} {91}},\ \bibinfo {pages} {014515} (\bibinfo {year}
  {2015})}\BibitemShut {NoStop}%
\bibitem [{\citenamefont {Baust}\ \emph {et~al.}(2016)\citenamefont {Baust},
  \citenamefont {Hoffmann}, \citenamefont {Haeberlein}, \citenamefont
  {Schwarz}, \citenamefont {Eder}, \citenamefont {Goetz}, \citenamefont
  {Wulschner}, \citenamefont {Xie}, \citenamefont {Zhong}, \citenamefont
  {Quijandr\'{\i}a}, \citenamefont {Zueco}, \citenamefont {Ripoll},
  \citenamefont {Garc\'{\i}a-\'Alvarez}, \citenamefont {Romero}, \citenamefont
  {Solano}, \citenamefont {Fedorov}, \citenamefont {Menzel}, \citenamefont
  {Deppe}, \citenamefont {Marx},\ and\ \citenamefont {Gross}}]{Baust_2016}%
  \BibitemOpen
  \bibfield  {author} {\bibinfo {author} {\bibfnamefont {A.}~\bibnamefont
  {Baust}}, \bibinfo {author} {\bibfnamefont {E.}~\bibnamefont {Hoffmann}},
  \bibinfo {author} {\bibfnamefont {M.}~\bibnamefont {Haeberlein}}, \bibinfo
  {author} {\bibfnamefont {M.~J.}\ \bibnamefont {Schwarz}}, \bibinfo {author}
  {\bibfnamefont {P.}~\bibnamefont {Eder}}, \bibinfo {author} {\bibfnamefont
  {J.}~\bibnamefont {Goetz}}, \bibinfo {author} {\bibfnamefont
  {F.}~\bibnamefont {Wulschner}}, \bibinfo {author} {\bibfnamefont
  {E.}~\bibnamefont {Xie}}, \bibinfo {author} {\bibfnamefont {L.}~\bibnamefont
  {Zhong}}, \bibinfo {author} {\bibfnamefont {F.}~\bibnamefont
  {Quijandr\'{\i}a}}, \bibinfo {author} {\bibfnamefont {D.}~\bibnamefont
  {Zueco}}, \bibinfo {author} {\bibfnamefont {J.-J.~G.}\ \bibnamefont
  {Ripoll}}, \bibinfo {author} {\bibfnamefont {L.}~\bibnamefont
  {Garc\'{\i}a-\'Alvarez}}, \bibinfo {author} {\bibfnamefont {G.}~\bibnamefont
  {Romero}}, \bibinfo {author} {\bibfnamefont {E.}~\bibnamefont {Solano}},
  \bibinfo {author} {\bibfnamefont {K.~G.}\ \bibnamefont {Fedorov}}, \bibinfo
  {author} {\bibfnamefont {E.~P.}\ \bibnamefont {Menzel}}, \bibinfo {author}
  {\bibfnamefont {F.}~\bibnamefont {Deppe}}, \bibinfo {author} {\bibfnamefont
  {A.}~\bibnamefont {Marx}}, \ and\ \bibinfo {author} {\bibfnamefont
  {R.}~\bibnamefont {Gross}},\ }\href {\doibase 10.1103/PhysRevB.93.214501}
  {\bibfield  {journal} {\bibinfo  {journal} {Phys.\,Rev.\,B}\ }\textbf
  {\bibinfo {volume} {93}},\ \bibinfo {pages} {214501} (\bibinfo {year}
  {2016})}\BibitemShut {NoStop}%
\bibitem [{\citenamefont {Braak}(2011)}]{Braak_2011}%
  \BibitemOpen
  \bibfield  {author} {\bibinfo {author} {\bibfnamefont {D.}~\bibnamefont
  {Braak}},\ }\href {\doibase 10.1103/PhysRevLett.107.100401} {\bibfield
  {journal} {\bibinfo  {journal} {Phys.\,Rev.\,Lett.}\ }\textbf {\bibinfo
  {volume} {107}},\ \bibinfo {pages} {100401} (\bibinfo {year}
  {2011})}\BibitemShut {NoStop}%
\bibitem [{\citenamefont {Pang}\ and\ \citenamefont
  {Jordan}(2017)}]{Pang_2017}%
  \BibitemOpen
  \bibfield  {author} {\bibinfo {author} {\bibfnamefont {S.}~\bibnamefont
  {Pang}}\ and\ \bibinfo {author} {\bibfnamefont {A.~N.}\ \bibnamefont
  {Jordan}},\ }\href {\doibase 10.1038/ncomms14695} {\bibfield  {journal}
  {\bibinfo  {journal} {Nat.\,Comm.}\ }\textbf {\bibinfo {volume} {8}},\
  \bibinfo {pages} {14695} (\bibinfo {year} {2017})}\BibitemShut {NoStop}%
\bibitem [{\citenamefont {Richer}\ and\ \citenamefont
  {DiVincenzo}(2016)}]{Richer_2016}%
  \BibitemOpen
  \bibfield  {author} {\bibinfo {author} {\bibfnamefont {S.}~\bibnamefont
  {Richer}}\ and\ \bibinfo {author} {\bibfnamefont {D.}~\bibnamefont
  {DiVincenzo}},\ }\href {\doibase 10.1103/PhysRevB.93.134501} {\bibfield
  {journal} {\bibinfo  {journal} {Phys.\,Rev.\,B}\ }\textbf {\bibinfo {volume}
  {93}},\ \bibinfo {pages} {134501} (\bibinfo {year} {2016})}\BibitemShut
  {NoStop}%
\bibitem [{\citenamefont {Porras}\ and\ \citenamefont
  {Garc\'{\i}a-Ripoll}(2012)}]{Porras_2012}%
  \BibitemOpen
  \bibfield  {author} {\bibinfo {author} {\bibfnamefont {D.}~\bibnamefont
  {Porras}}\ and\ \bibinfo {author} {\bibfnamefont {J.~J.}\ \bibnamefont
  {Garc\'{\i}a-Ripoll}},\ }\href {\doibase 10.1103/PhysRevLett.108.043602}
  {\bibfield  {journal} {\bibinfo  {journal} {Phys.\,Rev.\,Lett.}\ }\textbf
  {\bibinfo {volume} {108}},\ \bibinfo {pages} {043602} (\bibinfo {year}
  {2012})}\BibitemShut {NoStop}%
\bibitem [{\citenamefont {Quijandr\'{\i}a}\ \emph {et~al.}(2013)\citenamefont
  {Quijandr\'{\i}a}, \citenamefont {Porras}, \citenamefont
  {Garc\'{\i}a-Ripoll},\ and\ \citenamefont {Zueco}}]{Quijandria_2013}%
  \BibitemOpen
  \bibfield  {author} {\bibinfo {author} {\bibfnamefont {F.}~\bibnamefont
  {Quijandr\'{\i}a}}, \bibinfo {author} {\bibfnamefont {D.}~\bibnamefont
  {Porras}}, \bibinfo {author} {\bibfnamefont {J.~J.}\ \bibnamefont
  {Garc\'{\i}a-Ripoll}}, \ and\ \bibinfo {author} {\bibfnamefont
  {D.}~\bibnamefont {Zueco}},\ }\href {\doibase 10.1103/PhysRevLett.111.073602}
  {\bibfield  {journal} {\bibinfo  {journal} {Phys.\,Rev.\,Lett.}\ }\textbf
  {\bibinfo {volume} {111}},\ \bibinfo {pages} {073602} (\bibinfo {year}
  {2013})}\BibitemShut {NoStop}%
\bibitem [{\citenamefont {Zhao}\ \emph {et~al.}(2015)\citenamefont {Zhao},
  \citenamefont {Liu}, \citenamefont {Liu},\ and\ \citenamefont
  {Nori}}]{Zhao_2015}%
  \BibitemOpen
  \bibfield  {author} {\bibinfo {author} {\bibfnamefont {Y.-J.}\ \bibnamefont
  {Zhao}}, \bibinfo {author} {\bibfnamefont {Y.-L.}\ \bibnamefont {Liu}},
  \bibinfo {author} {\bibfnamefont {Y.-x.}\ \bibnamefont {Liu}}, \ and\
  \bibinfo {author} {\bibfnamefont {F.}~\bibnamefont {Nori}},\ }\href {\doibase
  10.1103/PhysRevA.91.053820} {\bibfield  {journal} {\bibinfo  {journal}
  {Phys.\,Rev.\,A}\ }\textbf {\bibinfo {volume} {91}},\ \bibinfo {pages}
  {053820} (\bibinfo {year} {2015})}\BibitemShut {NoStop}%
\bibitem [{\citenamefont {del Rey}\ \emph {et~al.}(2012)\citenamefont {del
  Rey}, \citenamefont {Porras},\ and\ \citenamefont
  {Martin-Martinez}}]{delRey_2012}%
  \BibitemOpen
  \bibfield  {author} {\bibinfo {author} {\bibfnamefont {M.}~\bibnamefont {del
  Rey}}, \bibinfo {author} {\bibfnamefont {D.}~\bibnamefont {Porras}}, \ and\
  \bibinfo {author} {\bibfnamefont {E.}~\bibnamefont {Martin-Martinez}},\
  }\href {\doibase 10.1103/PhysRevA.85.022511} {\bibfield  {journal} {\bibinfo
  {journal} {Phys.\,Rev.\,A}\ }\textbf {\bibinfo {volume} {85}},\ \bibinfo
  {pages} {022511} (\bibinfo {year} {2012})}\BibitemShut {NoStop}%
\bibitem [{\citenamefont {Abragam}(1961)}]{Abragam_1961}%
  \BibitemOpen
  \bibfield  {author} {\bibinfo {author} {\bibfnamefont {A.}~\bibnamefont
  {Abragam}},\ }\href@noop {} {\emph {\bibinfo {title} {The principles of
  nuclear magnetism}}},\ \bibinfo {number} {32}\ (\bibinfo  {publisher} {Oxford
  university press},\ \bibinfo {year} {1961})\BibitemShut {NoStop}%
\bibitem [{\citenamefont {Orlando}\ \emph {et~al.}(1999)\citenamefont
  {Orlando}, \citenamefont {Mooij}, \citenamefont {Tian}, \citenamefont
  {van~der Wal}, \citenamefont {Levitov}, \citenamefont {Lloyd},\ and\
  \citenamefont {Mazo}}]{Orlando_1999}%
  \BibitemOpen
  \bibfield  {author} {\bibinfo {author} {\bibfnamefont {T.~P.}\ \bibnamefont
  {Orlando}}, \bibinfo {author} {\bibfnamefont {J.~E.}\ \bibnamefont {Mooij}},
  \bibinfo {author} {\bibfnamefont {L.}~\bibnamefont {Tian}}, \bibinfo {author}
  {\bibfnamefont {C.~H.}\ \bibnamefont {van~der Wal}}, \bibinfo {author}
  {\bibfnamefont {L.~S.}\ \bibnamefont {Levitov}}, \bibinfo {author}
  {\bibfnamefont {S.}~\bibnamefont {Lloyd}}, \ and\ \bibinfo {author}
  {\bibfnamefont {J.~J.}\ \bibnamefont {Mazo}},\ }\href {\doibase
  10.1103/PhysRevB.60.15398} {\bibfield  {journal} {\bibinfo  {journal}
  {Phys.\,Rev.\,B}\ }\textbf {\bibinfo {volume} {60}},\ \bibinfo {pages}
  {15398} (\bibinfo {year} {1999})}\BibitemShut {NoStop}%
\bibitem [{\citenamefont {Schuster}\ \emph {et~al.}(2005)\citenamefont
  {Schuster}, \citenamefont {Wallraff}, \citenamefont {Blais}, \citenamefont
  {Frunzio}, \citenamefont {Huang}, \citenamefont {Majer}, \citenamefont
  {Girvin},\ and\ \citenamefont {Schoelkopf}}]{Schuster_2005}%
  \BibitemOpen
  \bibfield  {author} {\bibinfo {author} {\bibfnamefont {D.~I.}\ \bibnamefont
  {Schuster}}, \bibinfo {author} {\bibfnamefont {A.}~\bibnamefont {Wallraff}},
  \bibinfo {author} {\bibfnamefont {A.}~\bibnamefont {Blais}}, \bibinfo
  {author} {\bibfnamefont {L.}~\bibnamefont {Frunzio}}, \bibinfo {author}
  {\bibfnamefont {R.-S.}\ \bibnamefont {Huang}}, \bibinfo {author}
  {\bibfnamefont {J.}~\bibnamefont {Majer}}, \bibinfo {author} {\bibfnamefont
  {S.~M.}\ \bibnamefont {Girvin}}, \ and\ \bibinfo {author} {\bibfnamefont
  {R.~J.}\ \bibnamefont {Schoelkopf}},\ }\href {\doibase
  10.1103/PhysRevLett.94.123602} {\bibfield  {journal} {\bibinfo  {journal}
  {Phys.\,Rev.\,Lett.}\ }\textbf {\bibinfo {volume} {94}},\ \bibinfo {pages}
  {123602} (\bibinfo {year} {2005})}\BibitemShut {NoStop}%
\bibitem [{\citenamefont {Raab}\ and\ \citenamefont
  {De~Lange}(2005)}]{Raab_2005}%
  \BibitemOpen
  \bibfield  {author} {\bibinfo {author} {\bibfnamefont {R.~E.}\ \bibnamefont
  {Raab}}\ and\ \bibinfo {author} {\bibfnamefont {O.~L.}\ \bibnamefont
  {De~Lange}},\ }\href@noop {} {\emph {\bibinfo {title} {Multipole theory in
  electromagnetism: classical, quantum, and symmetry aspects, with
  applications}}},\ Vol.\ \bibinfo {volume} {128}\ (\bibinfo  {publisher}
  {Oxford University Press on Demand},\ \bibinfo {year} {2005})\BibitemShut
  {NoStop}%
\bibitem [{\citenamefont {Cohen-Tannoudji}\ \emph {et~al.}(2006)\citenamefont
  {Cohen-Tannoudji}, \citenamefont {Diu},\ and\ \citenamefont
  {Laloe}}]{Cohen_2006}%
  \BibitemOpen
  \bibfield  {author} {\bibinfo {author} {\bibfnamefont {C.}~\bibnamefont
  {Cohen-Tannoudji}}, \bibinfo {author} {\bibfnamefont {B.}~\bibnamefont
  {Diu}}, \ and\ \bibinfo {author} {\bibfnamefont {F.}~\bibnamefont {Laloe}},\
  }\href@noop {} {\emph {\bibinfo {title} {Quantum Mechanics}}},\ \bibinfo
  {number} {2}\ (\bibinfo  {publisher} {John Wiley \& Sons},\ \bibinfo {year}
  {2006})\BibitemShut {NoStop}%
\bibitem [{\citenamefont {Gambetta}\ \emph {et~al.}(2008)\citenamefont
  {Gambetta}, \citenamefont {Blais}, \citenamefont {Boissonneault},
  \citenamefont {Houck}, \citenamefont {Schuster},\ and\ \citenamefont
  {Girvin}}]{Gambetta_2008}%
  \BibitemOpen
  \bibfield  {author} {\bibinfo {author} {\bibfnamefont {J.}~\bibnamefont
  {Gambetta}}, \bibinfo {author} {\bibfnamefont {A.}~\bibnamefont {Blais}},
  \bibinfo {author} {\bibfnamefont {M.}~\bibnamefont {Boissonneault}}, \bibinfo
  {author} {\bibfnamefont {A.~A.}\ \bibnamefont {Houck}}, \bibinfo {author}
  {\bibfnamefont {D.~I.}\ \bibnamefont {Schuster}}, \ and\ \bibinfo {author}
  {\bibfnamefont {S.~M.}\ \bibnamefont {Girvin}},\ }\href {\doibase
  10.1103/PhysRevA.77.012112} {\bibfield  {journal} {\bibinfo  {journal}
  {Phys.\,Rev.\,A}\ }\textbf {\bibinfo {volume} {77}},\ \bibinfo {pages}
  {012112} (\bibinfo {year} {2008})}\BibitemShut {NoStop}%
\bibitem [{\citenamefont {Boissonneault}\ \emph {et~al.}(2009)\citenamefont
  {Boissonneault}, \citenamefont {Gambetta},\ and\ \citenamefont
  {Blais}}]{Boissonneault_2009}%
  \BibitemOpen
  \bibfield  {author} {\bibinfo {author} {\bibfnamefont {M.}~\bibnamefont
  {Boissonneault}}, \bibinfo {author} {\bibfnamefont {J.~M.}\ \bibnamefont
  {Gambetta}}, \ and\ \bibinfo {author} {\bibfnamefont {A.}~\bibnamefont
  {Blais}},\ }\href {\doibase 10.1103/PhysRevA.79.013819} {\bibfield  {journal}
  {\bibinfo  {journal} {Phys.\,Rev.\,A}\ }\textbf {\bibinfo {volume} {79}},\
  \bibinfo {pages} {013819} (\bibinfo {year} {2009})}\BibitemShut {NoStop}%
\bibitem [{\citenamefont {Niemczyk}\ \emph {et~al.}(2010)\citenamefont
  {Niemczyk}, \citenamefont {Deppe}, \citenamefont {H{\"u}bl}, \citenamefont
  {Menzel}, \citenamefont {Hocke}, \citenamefont {Schwarz}, \citenamefont
  {Garcia-Ripoll}, \citenamefont {Zueco}, \citenamefont {H{\"u}mmer},
  \citenamefont {Solano}, \citenamefont {Marx},\ and\ \citenamefont
  {Gross}}]{Niemczyk_2010}%
  \BibitemOpen
  \bibfield  {author} {\bibinfo {author} {\bibfnamefont {T.}~\bibnamefont
  {Niemczyk}}, \bibinfo {author} {\bibfnamefont {F.}~\bibnamefont {Deppe}},
  \bibinfo {author} {\bibfnamefont {H.}~\bibnamefont {H{\"u}bl}}, \bibinfo
  {author} {\bibfnamefont {E.~P.}\ \bibnamefont {Menzel}}, \bibinfo {author}
  {\bibfnamefont {F.}~\bibnamefont {Hocke}}, \bibinfo {author} {\bibfnamefont
  {M.~J.}\ \bibnamefont {Schwarz}}, \bibinfo {author} {\bibfnamefont {J.~J.}\
  \bibnamefont {Garcia-Ripoll}}, \bibinfo {author} {\bibfnamefont
  {D.}~\bibnamefont {Zueco}}, \bibinfo {author} {\bibfnamefont
  {T.}~\bibnamefont {H{\"u}mmer}}, \bibinfo {author} {\bibfnamefont
  {E.}~\bibnamefont {Solano}}, \bibinfo {author} {\bibfnamefont
  {A.}~\bibnamefont {Marx}}, \ and\ \bibinfo {author} {\bibfnamefont
  {R.}~\bibnamefont {Gross}},\ }\href {\doibase 10.1038/nphys1730} {\bibfield
  {journal} {\bibinfo  {journal} {Nat.\,Phys.}\ }\textbf {\bibinfo {volume}
  {6}},\ \bibinfo {pages} {772} (\bibinfo {year} {2010})}\BibitemShut {NoStop}%
\bibitem [{\citenamefont {{Niemczyk}}\ \emph {et~al.}(2011)\citenamefont
  {{Niemczyk}}, \citenamefont {{Deppe}}, \citenamefont {{Menzel}},
  \citenamefont {{Schwarz}}, \citenamefont {{Huebl}}, \citenamefont {{Hocke}},
  \citenamefont {{H{\"a}berlein}}, \citenamefont {{Danner}}, \citenamefont
  {{Hoffmann}}, \citenamefont {{Baust}}, \citenamefont {{Solano}},
  \citenamefont {{Garcia-Ripoll}}, \citenamefont {{Marx}},\ and\ \citenamefont
  {{Gross}}}]{Niemczyk_2011}%
  \BibitemOpen
  \bibfield  {author} {\bibinfo {author} {\bibfnamefont {T.}~\bibnamefont
  {{Niemczyk}}}, \bibinfo {author} {\bibfnamefont {F.}~\bibnamefont {{Deppe}}},
  \bibinfo {author} {\bibfnamefont {E.~P.}\ \bibnamefont {{Menzel}}}, \bibinfo
  {author} {\bibfnamefont {M.~J.}\ \bibnamefont {{Schwarz}}}, \bibinfo {author}
  {\bibfnamefont {H.}~\bibnamefont {{Huebl}}}, \bibinfo {author} {\bibfnamefont
  {F.}~\bibnamefont {{Hocke}}}, \bibinfo {author} {\bibfnamefont
  {M.}~\bibnamefont {{H{\"a}berlein}}}, \bibinfo {author} {\bibfnamefont
  {M.}~\bibnamefont {{Danner}}}, \bibinfo {author} {\bibfnamefont
  {E.}~\bibnamefont {{Hoffmann}}}, \bibinfo {author} {\bibfnamefont
  {A.}~\bibnamefont {{Baust}}}, \bibinfo {author} {\bibfnamefont
  {E.}~\bibnamefont {{Solano}}}, \bibinfo {author} {\bibfnamefont {J.~J.}\
  \bibnamefont {{Garcia-Ripoll}}}, \bibinfo {author} {\bibfnamefont
  {A.}~\bibnamefont {{Marx}}}, \ and\ \bibinfo {author} {\bibfnamefont
  {R.}~\bibnamefont {{Gross}}},\ }\href@noop {} {\bibfield  {journal} {\bibinfo
   {journal} {ArXiv}\ } (\bibinfo {year} {2011})},\ \Eprint
  {http://arxiv.org/abs/1107.0810} {arXiv:1107.0810 [cond-mat.mes-hall]}
  \BibitemShut {NoStop}%
\bibitem [{\citenamefont {Kasperczyk}\ \emph {et~al.}(2015)\citenamefont
  {Kasperczyk}, \citenamefont {Person}, \citenamefont {Ananias}, \citenamefont
  {Carlos},\ and\ \citenamefont {Novotny}}]{Kasperczyk_2015}%
  \BibitemOpen
  \bibfield  {author} {\bibinfo {author} {\bibfnamefont {M.}~\bibnamefont
  {Kasperczyk}}, \bibinfo {author} {\bibfnamefont {S.}~\bibnamefont {Person}},
  \bibinfo {author} {\bibfnamefont {D.}~\bibnamefont {Ananias}}, \bibinfo
  {author} {\bibfnamefont {L.~D.}\ \bibnamefont {Carlos}}, \ and\ \bibinfo
  {author} {\bibfnamefont {L.}~\bibnamefont {Novotny}},\ }\href {\doibase
  10.1103/PhysRevLett.114.163903} {\bibfield  {journal} {\bibinfo  {journal}
  {Phys.\,Rev.\,Lett.}\ }\textbf {\bibinfo {volume} {114}},\ \bibinfo {pages}
  {163903} (\bibinfo {year} {2015})}\BibitemShut {NoStop}%
\bibitem [{\citenamefont {Goetz}\ \emph
  {et~al.}(2017{\natexlab{b}})\citenamefont {Goetz}, \citenamefont {Deppe},
  \citenamefont {Eder}, \citenamefont {Fischer}, \citenamefont {M{\"u}ting},
  \citenamefont {{Puertas Mart{\'i}nez}}, \citenamefont {Pogorzalek},
  \citenamefont {Wulschner}, \citenamefont {Xie}, \citenamefont {Fedorov},
  \citenamefont {Marx},\ and\ \citenamefont {Gross}}]{Goetz_2016b}%
  \BibitemOpen
  \bibfield  {author} {\bibinfo {author} {\bibfnamefont {J.}~\bibnamefont
  {Goetz}}, \bibinfo {author} {\bibfnamefont {F.}~\bibnamefont {Deppe}},
  \bibinfo {author} {\bibfnamefont {P.}~\bibnamefont {Eder}}, \bibinfo {author}
  {\bibfnamefont {M.}~\bibnamefont {Fischer}}, \bibinfo {author} {\bibfnamefont
  {M.}~\bibnamefont {M{\"u}ting}}, \bibinfo {author} {\bibfnamefont
  {J.}~\bibnamefont {{Puertas Mart{\'i}nez}}}, \bibinfo {author} {\bibfnamefont
  {S.}~\bibnamefont {Pogorzalek}}, \bibinfo {author} {\bibfnamefont
  {F.}~\bibnamefont {Wulschner}}, \bibinfo {author} {\bibfnamefont
  {E.}~\bibnamefont {Xie}}, \bibinfo {author} {\bibfnamefont {K.~G.}\
  \bibnamefont {Fedorov}}, \bibinfo {author} {\bibfnamefont {A.}~\bibnamefont
  {Marx}}, \ and\ \bibinfo {author} {\bibfnamefont {R.}~\bibnamefont {Gross}},\
  }\href {\doibase 10.1088/2058-9565/aa66e7} {\bibfield  {journal} {\bibinfo
  {journal} {Quant.\,Sci.\,Tech.}\ }\textbf {\bibinfo {volume} {2}},\ \bibinfo
  {pages} {025002} (\bibinfo {year} {2017}{\natexlab{b}})}\BibitemShut
  {NoStop}%
\end{thebibliography}%

%%%%%%%%%% Merge with supplemental materials %%%%%%%%%%
\newpage
\widetext
\begin{center}
\textbf{\large Supplemental Materials: Parity-engineered light-matter interaction}
\end{center}
%%%%%%%%%% Merge with supplemental materials %%%%%%%%%%
%%%%%%%%%% Prefix a "S" to all equations, figures, tables and reset the counter %%%%%%%%%%
\setcounter{equation}{0}
\setcounter{figure}{0}
\setcounter{table}{0}
\makeatletter
\renewcommand{\theequation}{S\arabic{equation}}
\renewcommand{\thefigure}{S\arabic{figure}}
\renewcommand{\bibnumfmt}[1]{[S#1]}

\section*{Introduction}
This document contains Supplementary Materials for the main text in the following order: (1) Supplementary Methods for sample fabrication and measurement techniques, (2) theoretical derivations of selection rules (SRs), Supplementary Figure\,\ref{fig:Fig_S01} and Figure\,\ref{fig:Fig_S09}.

\section{Supplementary Methods}
\label{sec:fab}

In this section, we present fabrication details, sample parameters, and introduce the measurement setup used in the main article.\medskip

\textbf{Sample layout} The sample chip is mounted in a gold plated sample box made from copper as shown in Fig.\,\ref{fig:Fig_S01}\,(a). To connect the sample to coaxial cables, we use CPW/SMA adapters. The heart of the sample is a tunable-gap gradiometric flux qubit~\cite{Paauw_2009,Schwarz_2013} that can be controlled by two on-chip antennas as shown in Fig.\,\ref{fig:Fig_S01}\,(b). The area of each gradiometer loop is $\mathcal{A}_{\mathrm{gr}}\,{=}\,\SI{20}{\micro\meter}\,{\times}\,\SI{20}{\micro\meter}$ and the DC~SCQUID loop area is $\mathcal{A}_{\mathrm{sq}}\,{=}\,\SI{5}{\micro\meter}\,{\times}\,\SI{12}{\micro\meter}$. All qubit lines are \SI{500}{\nano\meter} wide. The two larger Josephson junctions have overlap areas of $\SI{250}{\nano\meter}\,{\times}\,\SI{200}{\nano\meter}$ [cf.~Fig.\,\ref{fig:Fig_S01}\,(c)] and the DC~SCQUID junctions have areas of $\SI{150}{\nano\meter}\,{\times}\,\SI{200}{\nano\meter}$. The resonator has a length of approximately \SI{15.25}{\milli\meter} and interdigital coupling capacitors as shown in Fig.\,\ref{fig:Fig_S01}\,(d). Additionally, we use quadratic \SI{8}{\micro\meter} wide holes in the ground plane to trap possible magnetic flux vortices.\medskip

\textbf{Spin coating parameters} We use spin coating to apply electron beam resist to the substrate. Before the spin coating, we clean the \SI{525}{\micro\meter} thick Si/SiO$_{2}$ substrate with an oxygen plasma for five minutes to remove residual particles on the surface. We then place it for \SI{10}{\minute} on a hotplate at \SI{160}{\celsius} to evaporate possible humidity at the substrate surface. For spin coating of the lower, $\SI{680}{\nano\meter}$ thick, layer of PMMA/MA$\SI{33}{\percent}$ resist, we use a rotation rate of $2000\,\mathrm{rpm}$. We bake the resist at $\SI{160}{\celsius}$ for $\SI{10}{\minute}$. In a second step, we coat the sample with $\SI{70}{\nano\meter}$ of PMMA950K resist, spinning at a rate of $4000\,\mathrm{rpm}$. Afterwards, we again bake the resist at $\SI{160}{\celsius}$ for $\SI{10}{\minute}$.\medskip

\textbf{Evaporation parameters} For metallization, we evaporate a $\SI{40}{\nano\meter}$ thick Al film at an angle of $\SI{17}{\degree}$. This layer forms the bottom electrode of the Josephson junctions but also covers the complete CPW structure. We then oxidize the sample for \SI{3000}{\second} at an oxygen pressure of \SI{3.3e-4}{\milli\bar} aiming at a current density of approximately $\SI{500}{\ampere\per\square\centi\meter}$. In a subsequent step, a $\SI{70}{\nano\meter}$ thick Al film is evaporated at an angle of $\SI{-17}{\degree}$, creating the second electrode of the junction. To oxidize all surfaces without contamination from air, we oxidize the sample once more inside the evaporation chamber. Here, we oxidize for \SI{3000}{\second} at an oxygen pressure of \SI{7.3e-3}{\milli\bar}.\medskip

\textbf{EBL parameters} We fabricate the samples using electron beam lithography (EBL). The flux qubit sample is fabricated in one fabrication step from an Al/AlO$_{\mathrm{x}}$/Al sandwich structure, i.e., we do not use Nb for the CPW structure. That way, we omit the lossy Al/Nb interface to increase the coherence times of qubit and resonator~\cite{Goetz_2016}. To pattern the structure in one single EBL process step, we use two independent electron beam currents at an acceleration voltage of \SI{80}{\kilo\volt}. In the first step, we pattern the large CPW structures and in the second step, we pattern the qubit and the Josephson junctions. For both steps, we use the auto conjugate function of a nB5 electron beam lithography system using gold nanoparticles to focus the beam. For the ground plane structures, we use a beam current of \SI{14.5}{\nano\ampere} and a dose of \SI{400}{\micro\coulomb\per\square\centi\meter}, while the center conductor and the antenna lines are written with a dose of \SI{480}{\micro\coulomb\per\square\centi\meter}. Furthermore, we scale the main-fields with a factor 1.001 to avoid stitching errors. The qubit is written with a beam current of \SI{2.6}{\nano\ampere} and a dose of \SI{800}{\micro\coulomb\per\square\centi\meter}. We develop the sample in two steps. First, we develop both resist layers using an AR600-56 developer for $\SI{45}{\second}$. Then, we immerse the sample in isopropanol at $\SI{4}{\celsius}$ for $\SI{120}{\second}$ to selectively develop only the lower resist layer. \medskip

\textbf{Cryogenic sample setup} For our experiments, we fix the sample with silver glue inside a gold-plated copper box and mount it to the base temperature of \SI{25}{\milli\kelvin} of our dilution refrigerator, i.e., \SI{25}{\milli\kelvin} [see Fig.\,\ref{fig:Fig_S01}\,(f)]. Our low temperature setup has a multistage shielding against magnetic flux noise containing $\hat{\mu}$-metal shields at room temperature, a cryoperm shield at \SI{4.2}{\kelvin} and an Al shield at the sample stage. Additionally, we use a gridded ground plane for the CPW resonator to prevent the motion of flux vortices.\medskip

\textbf{Measurement setup} The sample is connected to several RF and DC control lines as depicted in Fig.\,\ref{fig:Fig_S01}\,(g). We perform measurements mainly with a vector network analyzer (VNA) and characterize the qubit decay rate with a time-domain measurement setup. To control the amplitude distribution of the qubit drive, we use a single microwave source and a room temperature phase shifter to generate the phase shift $\varphi$ in one of the RF lines. By carefully equalizing the effective length of both antenna lines, we assure that there is no frequency dependent phase shift. Furthermore, we adjust the attenuation of the lines to generate drive fields with equal amplitudes. We filter input and output lines with band-pass filters at room temperature and at the sample stage to protect the qubit from RF noise entering the sample box through the resonator ports. That way, we further suppress the noise for frequencies outside the band-pass region. We control the qubit using two on-chip antennas, which are connected to bias tees for AC and DC control signals. These bias tees are specified to work between \SI{2}{\giga\hertz} and \SI{18}{\giga\hertz} for the RF input and have a \SI{200}{\mega\hertz} bandwidth of the DC port. Hence, the DC cables to the bias tees are additionally low-pass filtered at cryogenic temperatures. Using the DC port of these bias tees, we generate on-chip static magnetic fields with a controlled spatial amplitude distribution. Additionally, we use the on-chip antennas to generate an RF field with varying spatial distribution. We use a room temperature phase shifter to control the phase difference $\varphi$ between the two antenna lines. By measuring the transmitted power between the antenna lines and the resonator, we find a precise sinusoidal behavior of the field amplitude. From further auxiliary measurements, we find a flat frequency distribution in the relevant frequency regime above \SI{8}{\giga\hertz}, which ensures that the amplitude distribution of the RF field is not affected by the usage of additional microwave components such as mixers, bias tees or, filters. When carefully adjusting the attenuation of the input lines, we can generate a situation where the effective field across the qubit area vanishes. We optimize the amplitude distribution of the driving field by using an attenuator configuration that creates a maximum signal difference for a phase shift of $\pi$. We calibrate the coupling strength for both antenna lines individually by measuring the transmission from each antenna to the output of the resonator. We find an imbalance of \SI{14}{\decibel} between the two transmission signals, which we attribute to differences in the line attenuation, in on-chip interference effects, and in imperfections in the qubit geometry. Adjusting for this imperfection, we achieve an on/off ratio of \SI{30}{\decibel} between the situations of constructive and destructive interference of the antenna signals.\medskip

\textbf{Resonator characterization} In the following paragraphs, we characterize the sample using spectroscopic and time-resolved measurements. Spectroscopic measurements of qubit-resonator systems are a convenient method to access the excitation spectrum of the system~\cite{Schuster_2005,Niemczyk_2010}. With our sample architecture, we can either populate the resonator with readout photons through the input line or excite the qubit through the antenna lines. We first characterize the resonator with a single-tone transmission experiment to extract the resonator frequency $\omega_{\mathrm{r}}/2\pi\,{=}\,\SI{3.88}{\giga\hertz}$ and the total loss rate $\kappa_{\mathrm{tot}}/2\pi\,{=}\,\SI{2.5}{\mega\hertz}$. This loss rate is obtained from a Lorentzian fit to the resonator transmission spectrum and corresponds to a loaded quality factor $Q_{\ell}\,{\simeq}\,1500$. The resonator is limited by its external loss rate $\kappa_{\mathrm{x}}/2\pi\,{=}\,\SI{2.43}{\mega\hertz}$ and has an internal loss rate $\kappa_{\mathrm{i}}/2\pi\,{\simeq}\,\SI{70}{\kilo\hertz}$.\medskip

\textbf{Magnetic flux control of the qubit} Even though we shield the sample against external flux noise, we generate a static magnetic field using a superconducting coil attached to the sample holder to flux-bias the qubit. For the homemade coil with 1200 windings, we use \SI{100}{\micro\meter} thick NbTi wire embedded in a phosphor bronze matrix. Because the coil dimensions are much larger than the qubit area, we assume that the generated field is spatially homogeneous across the qubit area. Hence, one expects  the transition frequency of the flux qubit to only depend on the magnetic flux through the $\alpha$-SQUID. Due to a finite gradiometer quality~\cite{Schwarz_2013}, there is also a flux difference between the two gradiometer loops. This flux difference between the gradiometer loops arises mostly from the current circulating in the $\alpha$-SQUID, which converts the homogeneous field to an effective field gradient. To characterize the flux-dependent transition frequency of the qubit, we vary the magnetic field with with the on-chip antenna lines in order to identify qubit-resonator anticrossings. The anticrossings are located at $\Phi_{\mathrm{gr}}\,{=}\,N\Phi_{0}$, where $\Phi_{\mathrm{gr}}\,{=}\,\Phi_{\varepsilon1}\,{-}\,\Phi_{\varepsilon2}$ is the flux difference in the two gradiometer loops. From the spacing of the anticrossings, we calculate an effective mutual inductance $M_{\varepsilon}\,{\simeq}\,0.625\,\Phi_{0}/$mA$\,{=}\,\SI{1.25}{\pico\henry}$, which describes the field difference between the two gradiometer loops induced by the external coil.\medskip

\textbf{Qubit transition frequency} For suitable coil currents, we observe clear anticrossings between qubit and resonator frequency and extract the transverse coupling strength $g_{\mathrm{t}}/2\pi\,{\simeq}\,\SI{40}{\mega\hertz}$. To further characterize the qubit, we perform qubit spectroscopy by applying an additional RF tone with frequency $\omega$ in a two-tone experiment as shown in Fig.\,2\,(b) of the main text. For simplicity, we use only one of the antenna lines for the RF drive tone for these characterization measurements. In the two-tone experiment, we utilize the qubit-state-dependent AC Stark shift in the dispersive regime $(g_{\mathrm{t}}^{2}/\delta^{2}\,{\simeq}\,\num{e-4})$, where $\delta\,{\equiv}\,\omega_{\mathrm{q}}\,{-}\,\omega_{\mathrm{r}}$. To control the operating point, we use a DC flux-bias generated by the external coil as well as a local magnetic field generated by the on-chip antenna lines. The on-chip control is particularly important to adjust the magnetic energy bias $\varepsilon$, i.e., the tilt of the double well potential of the qubit. By adjusting $\varepsilon\,{\simeq}\,0$, we determine the qubit gap $\Delta$ from the center frequency of a Lorentzian fit. We perform additional qubit spectroscopy measurements at different operating points and we observe qubit gap frequencies up to a maximum $\Delta_{\mathrm{max}}/\hbar\,{\simeq}\,2\pi\,{\times}\,\SI{10.5}{\giga\hertz}$.\medskip

\textbf{Readout photons} Due to uncertainties in the transmission properties of both our measurement lines and the insertion loss of the resonator itself, we require an in situ calibration of the resonator population. To this end, we determine the readout photon number $\bar{n}\,{=}\,\langle\hat{a}^{\dagger}\hat{a}\rangle$, using the photon number dependence of the qubit frequency~\cite{Schuster_2005,Goetz_2016b}. We control $\bar{n}$ by varying the output power $\mathcal{P}_{\mathrm{r}}$ of the VNA and measure the frequency shift $\delta\omega_{\mathrm{q}}\,{=}\,2\bar{n}g_{\mathrm{t}}^{2}/\delta$ relative to the bare qubit transition frequency $\omega_{\mathrm{q},0}$. From the linear fit, we calculate a resonator population of 331\,photons/mW power emitted from the VNA. All measurements presented in the main article are performed with $\bar{n}\,{=}\,33$ photons on average. This value is still well below the critical photon number~\cite{Boissonneault_2009} $n_{\mathrm{crit}}\,{\equiv}\,\delta^{2}/(2g_{\mathrm{t}})^{2}\,{\simeq}\,1900$ above which the dispersive limit breaks down. The finite number of readout photons, however, increases the qubit dephasing rate accordingly due to measurement induced dephasing~\cite{Gambetta_2008,Goetz_2016a}.\medskip

\textbf{Qubit spectroscopy power} We calibrate the number of drive photons $\bar{n}_{\mathrm{d}}$ that arrive at the qubit for a given source power $\mathcal{P}_{\mathrm{d}}$ used for the drive tone. Similar to the AC Stark calibration used to calibrate $\bar{n}$, we calibrate the drive photons by detecting the power broadening~\cite{Abragam_1961,Schuster_2005} of the qubit linewidth
\begin{equation}
 \gamma_{\mathrm{q}} = \sqrt{\gamma_{2}^{2} + \bar{n}_{\mathrm{d}}(2g)^{2}\frac{\gamma_{2}}{\gamma_{1}} }\,.
\label{eqn:powerbroad}
\end{equation}
We extract the photon number by assuming a linear power-to-photon conversion $\alpha_{\mathrm{d}}$, i.e., $\bar{n}_{\mathrm{d}}\,{=}\,\alpha_{\mathrm{d}}\mathcal{P}_{\mathrm{s}}$. Using $\gamma_{2}/2\pi\,{=}\,\SI{9.7}{\mega\hertz}$ and $\gamma_{1}/2\pi\,{=}\,\SI{385}{\kilo\hertz}$ measured above, we obtain $\alpha_{\mathrm{d}}\,{\simeq}\,0.16$ photons/mW.

\section{Multipole expansion}
\label{sec:selrules}

\textbf{Dipole and Quadrupole moments} Here, we derive a multipole expansion~\cite{Kasperczyk_2015} for the coupling between a magnetic field $\mathbf{B}(\omega,t)\,{=}\,\mathbf{B}\cos\omega t$ and the tunable-gap gradiometric flux qubit. Here, $\omega$ is the drive frequency and $t$ is the time. Because the qubit is located in the $xy$-plane, only the $z$-component $B_{z}$ is relevant. To calculate the multipole coupling to this field component, we first calculate the dipole moment for a quasiplanar loop and the quadrupole moment for a gradiometer. For a quasiplanar loop with area $\mathcal{A}$ carrying a constant current $I$, the magnetic dipole moment~\cite{Raab_2005} $\mathbf{p}\,{=}\,(1/2)\int_{\mathcal{V}}\mathrm{d}^{3}r\,\mathbf{r}\,{\times}\,\mathbf{j}$ can be simplified to $\mathbf{p}\,{=}\,(I/2)\int_{\partial\mathcal{A}}\mathbf{r}\,{\times}\,\mathrm{d}\mathbf{r}\,{=}\,(0,0,I\mathcal{A})$. This leads to the dipole moment $p\,{=}\,|\mathbf{p}|\,{=}\,|I|\mathcal{A}$ stated in the main text where the area is the SQUID area, $\mathcal{A}\,{=}\,\mathcal{A}_{\mathrm{sq}}$. For the components of the magnetic quadrupole moment we obtain $Q_{ij}\,{=}\,(2I/3)\sum_{k}\mathbf{p}_{k,i}r_{k,j}$, where $\mathbf{p}_{k}$ is the $k^{\mathrm{th}}$ dipole positioned at $\mathbf{r}_{k}$. For a single loop $(k\,{=}\,1)$ and $\mathbf{r}_{1}\,{=}\,(0,0,0)$, all components of $\textbf{Q}$ are zero. For the quadrupolar gradiometer case in the main text, the dipole moment is zero and $\textbf{Q}$ has the finite component $Q\,{\equiv}\,Q_{xz}\,{=}\,4I\mathcal{A}d/3$, where $\mathcal{A}\,{=}\,\mathcal{A}_{\mathrm{gr}}$ is the area of a single gradiometer loop.\medskip

\textbf{Single-loop flux qubit} Integrating three Josephson junctions with suitable parameters into a single superconducting loop forms a flux qubit with persistent current $I_{\mathrm{p}}$ and area $\mathcal{A}_{\mathrm{q}}$. In the bare qubit basis, the dipolar interaction Hamiltonian between the qubit and an oscillating magnetic field reads $\widehat{\mathcal{H}}_{\mathrm{int}}^{p}\,{=}\,B_{z0}p\hat{\sigma}_{x}$. Here, $\mathbf{p}\,{=}\,I_{\mathrm{p}}\mathcal{A}_{\mathrm{q}}$ is the qubit dipole moment and $B_{z0}\,{=}\,\mathcal{A}_{\mathrm{q}}^{-1}\int_{\mathcal{A}_{\mathrm{q}}}B_{z}(\mathbf{r})\mathrm{d}^{2}r$ is the effective magnetic field penetrating the qubit loop. When exposing the qubit to a constant field gradient in $x$-direction with $B_{z}(L/2)\,{=}\,0$, the interaction strength vanishes. For arbitrary field gradients, the interaction Hamiltonian reads $\widehat{\mathcal{H}}_{\mathrm{int}}^{Q}\,{=}\,Q(\delta B_{z}/\delta x)\hat{\sigma}_{x}$, where $(\delta B_{z}/\delta x)\,{=}\,\mathcal{A}_{\mathrm{q}}^{-1}\int_{\mathcal{A}_{\mathrm{q}}}\partial_{x}B_{z}(\mathbf{r})\mathrm{d}^{2}r$ is the effective magnetic field gradient penetrating the qubit loop.\medskip

\textbf{Gap-tunable gradiometric flux qubit} We now turn to the special case of gradiometric flux qubits with tunable gap, where a DC~SQUID replaces the $\alpha$ junction on the center line of a gradiometer structure~\cite{Paauw_2009,Schwarz_2013}. Here, we only consider a symmetric situation, where two currents split equally in the two gradiometer parts and flow in opposite direction on the center line including the Josephson junctions. Then, antisymmetric fields create transversal coupling via the quadrupole moment $Q$. In the ideal case, the two currents cancel each other on the center line and the SQUID dipole moment $p$ does not couple to the qubit phase $\theta_{\mathrm{q}}$ and is thus irrelevant for the qubit. However, because the flux threading the SQUID loop changes the qubit gap $\Delta$ the symmetric part of the magnetic field creates longitudinal coupling proportional to the SQUID dipole moment $p$. This situation leads to the Hamiltonian $\widehat{\mathcal{H}}$ presented in the main text.

\section{Selection rules}
\label{sec:selrules}

A selection rule constrains the possible transition between two quantum states induced by an external drive due to the conservation of different quantum numbers such as angular momentum or parity~\cite{Cohen_2006}. Originating from quantum optics, SRs are also valid for circuit QED experiments using flux qubits as artificial atoms~\cite{Liu_2005,Deppe_2008,Niemczyk_2011,Forn-Diaz_2015}. In our case, the corresponding quantum states are either reflected by the initial and final state of the qubit, or by the dressed qubit-resonator states, if we are probing sideband transitions. To derive SRs, we calculate the transition moment integral$\int\psi_{\mathrm{f}}\hat{\mu}\psi_{\mathrm{i}}$. Here, $\psi_{\mathrm{f}}$ and $\psi_{\mathrm{i}}$ are the wave functions of final and initial state and $\hat{\mu}$ is the interaction operator. The terminology of which transitions are called ``allowed'' and which ones are called ``forbidden'' is based on a hydrogen-like atom with 1s1 electron configuration. For superconducting flux qubits, SRs reduce to pure parity arguments, and transitions are allowed if the total parity of $\psi_{\mathrm{f}}\hat{\mu}\psi_{\mathrm{i}}$ is even. Hence, an odd-parity $\hat{\mu}$ induces transitions between states of equal parity and an even-parity $\hat{\mu}$ induces transitions between states of equal parity. For atoms and qubits, $\hat{\mu}$ is represented by an external electromagnetic drive field. Concerning the parity of drive fields, one has to consider the structure of $E$- and $B$-fields in Maxwell's equations. Generally, one finds that the electric field has even parity and the magnetic field has odd parity. Consequently, an electric field gradient has even parity and a magnetic field gradient has odd parity. For qubits with corresponding parity operator $\widehat{\Pi}_{\mathrm{q}}\,{=}\,{-}\hat{\sigma}_{z}$, one finds that $\hat{\sigma}_{x}$ is odd and $\hat{\sigma}_{z}$ is even by calculating the commutator and anitcommutator relations~\cite{Deppe_2008}.\medskip

For flux qubits at the degeneracy point, the ground state $\ket{\mathrm{g}}$ has even parity, while the excited state $\ket{\mathrm{e}}$ has odd parity~\cite{Orlando_1999}. In the same way as for $\widehat{\Pi}_{\mathrm{q}}$, we define an even operator~$\hat{A}_{+}$ if it commutes with $\ket{\mathrm{g}}$ and vice versa. Even operators cannot induce transitions between states of different parities, which is expressed in the vanishing matrix element $\braket{\mathrm{e}|\hat{A}_{+}|\mathrm{g}}\,{=}\,0$. On the other hand, an odd operator $\hat{A}_{-}$ can induce transitions between $\ket{\mathrm{g}}$ and $\ket{\mathrm{e}}$, i.e.,$\braket{\mathrm{e}|\hat{A}_{-}|\mathrm{g}}\,{>}\,0$. Using this formalism, one finds that $\hat{\sigma}_{x}$ is an odd operator, while $\hat{\sigma}_{z}$ is even if the qubit states have opposite parity~\cite{Cohen_2006}. This fact results in dipolar SRs for $\hat{\sigma}_{x}$-interactions and quadrupolar SRs for $\hat{\sigma}_{z}$-interactions. If the two participating quantum states have equal parity, the situation changes and even operators induce transitions while they are forbidden for odd operators.\medskip

\begin{figure*}[h!]
 \centering
 \includegraphics{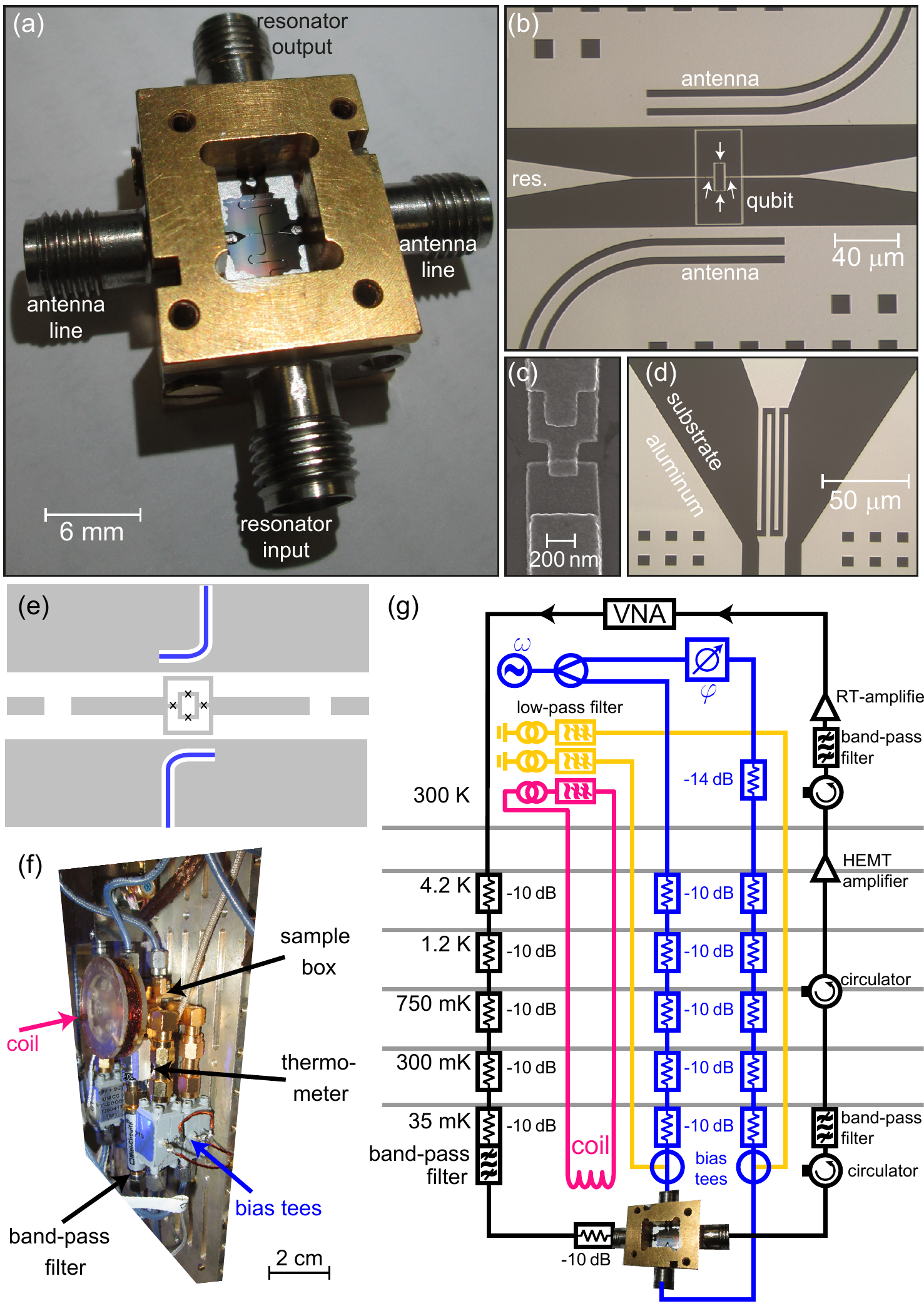}
 \caption{\textbf{(a)} Photograph of the sample box for the flux qubit sample including four SMA connectors for the RF-lines and the sample chip in the center. The resonator meanders across the substrate and at the corners of the substrate, silver-glue is visible. \textbf{(b)} Microscope image of the area where the tunable-gap gradiometric flux qubit is located. The resonator center conductor runs from left to right and is galvanically coupled to the qubit. There are two antenna lines approaching from the bottom and from the top to shape the driving field. Arrows indicate the positions where Josephson junctions are located. \textbf{(c)} Scanning electron microscope image showing one of the two larger Josephson junctions. \textbf{(d)} Microscope image of one of the coupling capacitors confining the half-wavelength resonator. \textbf{(e)} Sketch of a flux qubit galvanically coupled to a readout resonator and inductively coupled to two on-chip antennas. \textbf{(f)} Photograph of the cryogenic measurement setup. \textbf{(g)} Detailed measurement setup including microwave and DC components.}
 \label{fig:Fig_S01}
\end{figure*}

\textbf{One-photon transitions for a $\hat{\sigma}_{x}\,{+}\,\hat{\sigma}_{z}$ drive} For the tunable-gap gradiometric flux qubit placed between two antennas as depicted in Fig.\,\ref{fig:Fig_S01}\,(b), we can induce longitudinal interaction with a symmetric, in the ideal case spatially homogeneous, microwave drive $\mathbf{B}\,{=}\,(0,0,B_{z})$. Furthermore, we can induce transversal interaction with an antisymmetric microwave drive, i.e., a pure field gradient $\nabla\mathbf{B}\,{\propto}\,(0,0,x)$. These fields reflect the variables $B_{z0}$ and $\delta B_{z}/\delta x$ in the main article. Here, we describe the two drives in terms of their creation and annihilation operators $\hat{\ell}^{\dagger},\hat{\ell}$ (longitudinal drive) and $\hat{t}^{\dagger},\hat{t}$ (transversal drive), respectively. Hence, the operator $\hat{\ell}$ creates a symmetric field distribution, while $\hat{t}$ creates an antisymmetric field distribution. Both driving fields are assumed to be coherent states, $\ket{\beta_{\ell,\mathrm{t}}}$, where $\hat{\ell}\ket{\beta_{\ell}}\,{=}\,\beta_{\ell}\ket{\beta_{\ell}}$ and $\hat{t}\ket{\beta_{\mathrm{t}}}\,{=}\,\beta_{\mathrm{t}}\ket{\beta_{\mathrm{t}}}$. The respective interaction with the qubit in the bare basis can be expressed as
\begin{align}
 \widehat{\mathcal{H}}_{\mathrm{int,}\ell} &= \hbar g_{\ell,\mathrm{d}}e^{\imath\omega t}(\hat{\ell}^{\dagger}\,{+}\,\hat{\ell})\hat{\sigma}_{x} \overset{\text{CL}}{\approx} \hbar \frac{\Omega_{\ell}}{2}\cos(\omega t)\hat{\sigma}_{x}\,,\label{eqn:omegal}\\
 \widehat{\mathcal{H}}_{\mathrm{int,t}} &= \hbar g_{\mathrm{t,d}}e^{\imath\omega t}(\hat{t}^{\dagger}\,{+}\,\hat{t})\hat{\sigma}_{z} \overset{\text{CL}}{\approx} \hbar \frac{\Omega_{\mathrm{t}}}{2}\cos(\omega t)\hat{\sigma}_{z}\,.\label{eqn:omega t}
\end{align}
Here, $g_{\ell,\mathrm{d}}$ and $g_{\mathrm{t,d}}$ denote the longitudinal and transversal vacuum coupling strength between the microwave drive and the qubit, respectively. In the classical limit (CL) on the right hand side of Eq.\,(\ref{eqn:omegal}) and Eq.\,(\ref{eqn:omega t}), the drives with frequency $\omega$ are characterized by their amplitudes $\Omega_{\ell}\,{=}\,4\hbar g_{\ell,\mathrm{d}}\beta_{\ell}$ and $\Omega_{\mathrm{t}}\,{=}\,4\hbar g_{\mathrm{t,d}}\beta_{\mathrm{t}}$, respectively. For simplicity, we neglect cross-coupling due to imperfections of the qubit structure in the following. Then, applied to the eigenstates $\ket{\mathrm{e}}$ and $\ket{\mathrm{g}}$, one finds that Eq.\,(\ref{eqn:omegal}) anti-commutes with the qubit operator $\hat{\sigma}_{z}$, while Eq.\,(\ref{eqn:omega t}) commutes when operating at the qubit degeneracy point. For a superposition of both drives and away from the degeneracy point, however, the system parity is not well-defined and the transition probabilities change as discussed below.\smallskip

In the following calculations, we derive transition matrix elements for qubit transitions under a mixed $(\Omega_{\ell}\,{+}\,\Omega_{\mathrm{t}})$ drive. We show that the selection rules known for circuit QED setups~\cite{Liu_2005,Deppe_2008,Niemczyk_2011} must be modified taking the longitudinal drive into account. We start our calculations with the coupled qubit-resonator Hamiltonian without the external drive:
\begin{align}
 \widehat{\mathcal{H}}_{\mathrm{sys,b}} = &\frac{\hbar\Delta}{2}\hat{\sigma}_{x}+\frac{\hbar\varepsilon}{2}\hat{\sigma}_{z}+\hbar\omega_{\mathrm{r}}\hat{a}^{\dag}\hat{a} + \hbar g_{\mathrm{t}}(\hat{a}\,{+}\,\hat{a}^{\dag})\hat{\sigma}_{z} + \hbar g_{\ell}(\hat{a}+\hat{a}^{\dag})\hat{\sigma}_{x}\,.
\label{eq:Htg}
\end{align}
In a next step, we transfer Eq.\,(\ref{eq:Htg}) to the qubit eigenenergy basis, which yields 
\begin{align}
 \widehat{\mathcal{H}}_{\mathrm{sys,q}} = &\frac{\hbar\omega_{\mathrm{q}}}{2}\hat{\sigma}_{z}+\hbar\omega_{\mathrm{r}}\hat{a}^{\dag}\hat{a}\notag\\
 &+\hbar g_{\mathrm{t}}\cos\theta(\hat{a}+\hat{a}^{\dag})\hat{\sigma}_{z} -\hbar g_{\mathrm{t}}\sin\theta(\hat{a}+\hat{a}^{\dag})\hat{\sigma}_{x}\notag\\
 &+\hbar g_{\ell}\cos\theta(\hat{a}+\hat{a}^{\dag})\hat{\sigma}_{x}+\hbar g_{\ell}\sin\theta(\hat{a}+\hat{a}^{\dag})\hat{\sigma}_{z}\,.
\label{eq:Heigen}
\end{align}
Here, we introduce the Bloch angle~$\theta\,{=}\,\tan^{-1}(\Delta/\varepsilon)$ and~$\omega_{\mathrm{q}}\,{=}\,\sqrt{\Delta^{2}\,{+}\,\varepsilon^{2}}$. At the flux degeneracy point, we find $\theta\,{=}\,\pi/2$ and $\{\cos\theta,\,\sin\theta\}\,{=}\,\{0,\,1\}$.\\
With Eq.~(\ref{eq:Htg}) and  Eq.~(\ref{eq:Heigen}) we describe the qubit-resonator coupling in its most general form, which means there could be both longitudinal and transversal coupling at the sweet spot. The longitudinal coupling would be present if the resonator current induces flux into the DC SQUID. For our specific sample geometry, however, the longitudinal coupling strength $g_{\ell}$ between qubit and resonator vanishes. This is because we place the DC SQUID on the symmetry axis of the gradiometric qubit. Therefore, most current runs on the outer lines because the Josephson inductance on the center line and of the DC SQUID strongly damps any AC current. Even if a small residual current runs through the center line, it will to first order split symmetrically into both SQUID arms and therefore not couple any flux into the DC SQUID. Hence, there is no longitudinal coupling at the sweet spot for our geometry. We could introduce such a coupling mechanism by rotating the qubit by \SI{90}{\degree} with respect to the resonator. In this case, the resonator would induce flux into the DC SQUID. The transversal coupling is mainly determined by the flux difference that the resonator induces into the gradiometer loop. The resulting coupling strength $g_{\mathrm{t}}$ depends on the special geometry and the mutual inductance between the two gradiometer loops and the resonator. Hence, it can be considered constant for our sample.\smallskip

We now add the two drive terms defined in Eq.\,(\ref{eqn:omegal}) and Eq.\,(\ref{eqn:omega t}) to $\widehat{\mathcal{H}}_{\mathrm{sys,q}}$, which results in
\begin{align}
 \widehat{\mathcal{H}}_{\mathrm{tot}} &= \frac{\hbar\omega_{\mathrm{q}}}{2}\hat{\sigma}_{z}+\hbar\omega_{\mathrm{r}}\hat{a}^{\dag}\hat{a}\notag\\
 & +\hbar g_{\mathrm{t}}\cos\theta(\hat{a}+\hat{a}^{\dag})\hat{\sigma}_{z} -\hbar g_{\mathrm{t}}\sin\theta(\hat{a}+\hat{a}^{\dag})\hat{\sigma}_{x}\notag\\
 &+\hbar g_{\ell}\cos\theta(\hat{a}+\hat{a}^{\dag})\hat{\sigma}_{x}+\hbar g_{\ell}\sin\theta(\hat{a}+\hat{a}^{\dag})\hat{\sigma}_{z}\notag\\
 &+\frac{\hbar\Omega_{\mathrm{t}}}{2}\cos\theta\cos(\omega t)\hat{\sigma}_{z}-\frac{\hbar\Omega_{\mathrm{t}}}{2}\sin\theta\cos(\omega t)\hat{\sigma}_{x}\notag\\
 &+\frac{\hbar\Omega_{\ell}}{2}\cos\theta\cos(\omega t)\hat{\sigma}_{x}+\frac{\hbar\Omega_{\ell}}{2}\sin\theta\cos(\omega t)\hat{\sigma}_{z}\,.
\label{eq:Hlt,q}
\end{align}
Next, we cancel the time-dependent terms~$\frac{1}{2}\Omega_{\mathrm{t}}\cos\theta\cos(\omega t)\hat{\sigma}_{z}$ and~$\frac{1}{2}\Omega_{\ell}\sin\theta\cos(\omega t)\hat{\sigma}_{z}$ by moving to a nonuniformly rotating frame, where~${\widehat{\mathcal{H}}}_{\mathrm{rot}}\,{=}\,\widehat{\mathcal{U}}\widehat{\mathcal{H}}_{\mathrm{tot}}\widehat{\mathcal{U}}^{\dagger}\,{-}\,\imath\hbar\partial\widehat{\mathcal{U}}\partial\widehat{\mathcal{U}}^{\dagger}/\partial t$, and we chose
\begin{equation}
\widehat{\mathcal{U}} = \exp\left[\frac{\imath}{2}\hat{\sigma}_{z}\sin(\omega t)\left(\frac{\Omega_{\mathrm{t}}}{\omega}\cos\theta+\frac{\Omega_{\ell}}{\omega}\sin\theta\right)\right]\,.
\label{eq:U_non}
\end{equation}
That way, the effective Hamiltonian reads
\begin{align}
 \widehat{\mathcal{H}}_{\mathrm{rot}} &= \frac{\hbar\omega_{\mathrm{q}}}{2}\hat{\sigma}_{z}+\hbar\omega_{\mathrm{r}}\hat{a}^{\dag}\hat{a}+\hbar g_{\mathrm{t}}\cos\theta(\hat{a}+\hat{a}^{\dag})\hat{\sigma}_{z} +\hbar g_{\ell}\sin\theta(\hat{a}+\hat{a}^{\dag})\hat{\sigma}_{z}\notag\\
 &+\hbar\left[g_{\ell}\cos\theta(\hat{a}+\hat{a}^{\dag})-g_{\mathrm{t}}\sin\theta(\hat{a}+\hat{a}^{\dag})+\frac{\Omega_{\ell}}{2}\cos\theta\cos(\omega t)-\frac{\Omega_{\mathrm{t}}}{2}\sin\theta\cos(\omega t)\right]\notag\\
 &\times\left[\hat{\sigma}_{+}e^{-\imath\phi}+\hat{\sigma}_{-}e^{+\imath\phi}\right]\,,
\label{eq:Hlt_tilde}
\end{align}
where~$\phi\,{=}\,{-}\sin(\omega t)(\Omega_{\mathrm{t}}\cos\theta\,{+}\,\Omega_{\ell}\sin\theta)/\omega$. We now move to the interaction picture with respect to qubit and resonator, which yields in the rotating wave approximation
\begin{align}
 \widehat{\mathcal{H}}_{\mathrm{eff}} &= \hbar\left[g_{\ell}\cos\theta(\hat{a}e^{-\imath\omega_{\mathrm{r}}t}+\hat{a}^{\dag}e^{+\imath\omega_{\mathrm{r}}t})- g_{\mathrm{t}}\sin\theta(\hat{a}e^{-\imath\omega_{\mathrm{r}}t}+e^{+\imath\omega_{\mathrm{r}}t}\hat{a}^{\dag})\right.\notag\\
 &\left.+\frac{\Omega_{\ell}}{2}\cos\theta\cos(\omega t)-\frac{\Omega_{\mathrm{t}}}{2}\sin\theta\cos(\omega t)\right]\times\left[\hat{\sigma}_{+}e^{+\imath\omega_{\mathrm{q}}t}e^{-\imath\phi}+\hat{\sigma}_{-}e^{-\imath\omega_{\mathrm{q}}t}e^{+\imath\phi}\right]\,.\notag
\label{eq:Hlt_int}
\end{align}

\begin{figure}[t]
\centering
\includegraphics{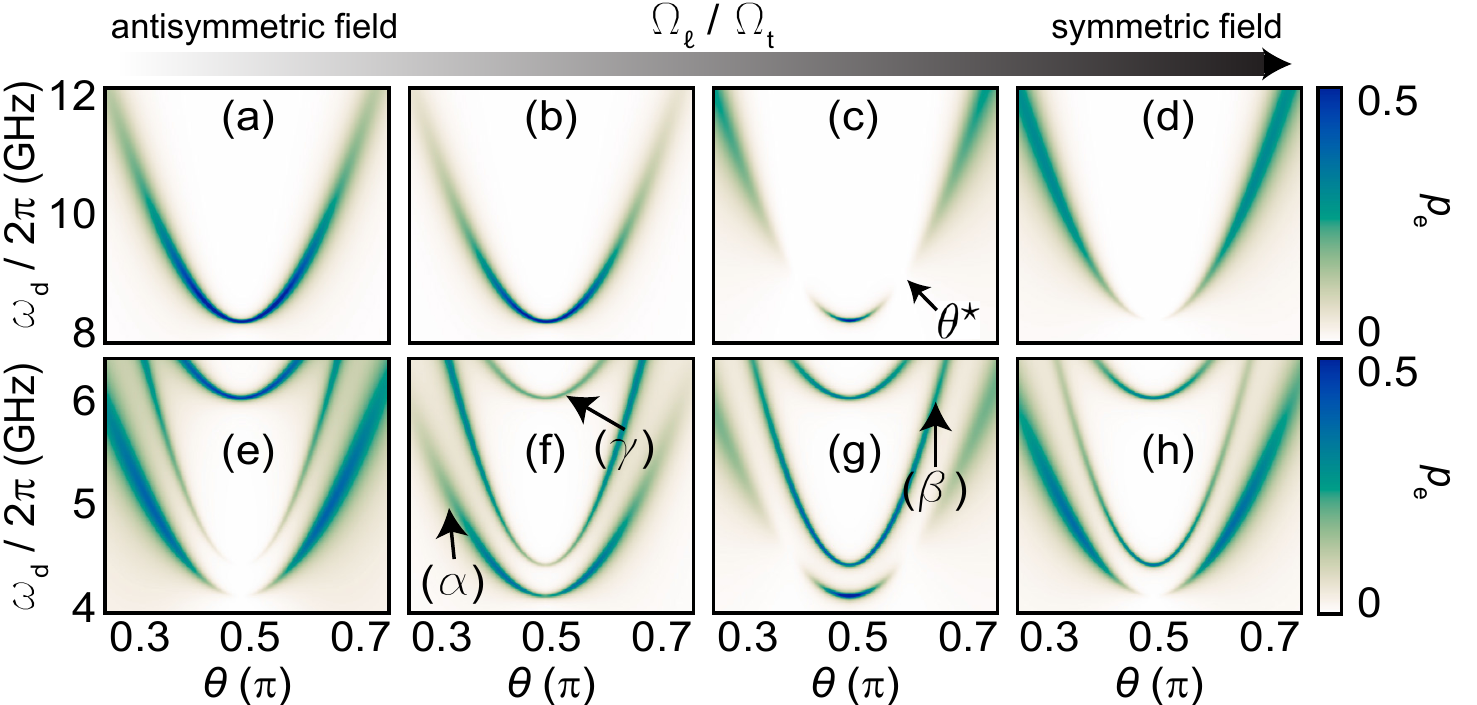}
\caption{Top row from \textbf{(a)} to \textbf{(d)}: Color encoded excitation spectra for the direct one-photon transition of a qubit under a continuous drive plotted versus Bloch angle $\theta$ and drive frequency calculated. The transition probabilities are calculated from the Bessel functions defined in Eq.\,(\ref{eq:H_tr1}). From panel (a) to panel (e) we increase the ratio~$\Omega_{\ell}/\Omega_{\mathrm{t}}$. To model the qubit linewidth, we superimpose the qubit hyperbola $\omega_{\mathrm{q}}\,{=}\,\hbar^{-1}\sqrt{\Delta^{2}\,{+}\,\varepsilon^{2}}$ with a Lorentzian lineshape, which increases proportional to $\gamma_{\mathrm{q}}(\theta)\,{\propto}\,\gamma_{2}\,{+}\,\gamma_{\phi}|\theta\,{-}\,\pi/2|$. In panel \textbf{(c)}, we show the situation $\Omega_{\ell}/\Omega_{\mathrm{t}}\,{=}\,\Delta/|\varepsilon|$  at the angle $\theta^{\star}$ where longitudinal coupling-induced transparency appears. Bottom row from \textbf{(e)} to \textbf{(h)}: Color encoded excitation spectra for the two-photon process $(\alpha)$, the red sideband $(\beta)$, and the two-photon process of the blue sideband $(\gamma)$ for a resonator frequency~$\omega_{\mathrm{r}}\,{=}\,\SI{4}{\giga\hertz}$.}
\label{fig:Fig_S09}
\end{figure}

We split this interaction Hamiltonian into a qubit-resonator term~$\widehat{\mathcal{H}}_{\mathrm{eff,r}}$ and into a qubit-driving term~$\widehat{\mathcal{H}}_{\mathrm{eff,q}}$ defined as
\begin{align}
 \widehat{\mathcal{H}}_{\mathrm{eff,r}} = \hbar&\left[g_{\ell}\cos\theta(\hat{a}e^{-\imath\omega_{\mathrm{r}}t}+\hat{a}^{\dag}e^{+\imath\omega_{\mathrm{r}}t})- g_{\mathrm{t}}\sin\theta(\hat{a}e^{-\imath\omega_{\mathrm{r}}t}+e^{+\imath\omega_{\mathrm{r}}t}\hat{a}^{\dag})\right]\notag\\
 \times&\left[\hat{\sigma}_{+}e^{+\imath\omega_{\mathrm{q}}t}e^{-\imath\phi}+\hat{\sigma}_{-}e^{-\imath\omega_{\mathrm{q}}t}e^{+\imath\phi}\right]\,,\\
 \widehat{\mathcal{H}}_{\mathrm{eff,q}} = \hbar&\left[\frac{\Omega_{\ell}}{2}\cos\theta-\frac{\Omega_{\mathrm{t}}}{2}\sin\theta\right]\cos(\omega t)\left[\hat{\sigma}_{+}e^{+\imath\omega_{\mathrm{q}}t}e^{-\imath\phi}+\hat{\sigma}_{-}e^{-\imath\omega_{\mathrm{q}}t}e^{+\imath\phi}\right]\,.
\label{eq:Hlt_eff}
\end{align}
For low power (one-photon) driving, the transition Hamiltonian can be approximated using Bessel functions, which results in
\begin{equation}
 \widehat{\mathcal{H}}_{\mathrm{trans},1} = \frac{\hbar}{2}\left[\frac{\Omega_{\ell}}{2}\cos\theta-\frac{\Omega_{\mathrm{t}}}{2}\sin\theta\right][J_{0}(\lambda)+J_{2}(\lambda)]\hat{\sigma}_{x} \approx\frac{\hbar}{2}\left[\frac{\Omega_{\ell}}{2}\cos\theta-\frac{\Omega_{\mathrm{t}}}{2}\sin\theta\right]\hat{\sigma}_{x} \,.
\label{eq:H_tr1}
\end{equation}
Here, $J_k$ is the~$k^{\mathrm{th}}$ Bessel function of the first kind and~$\lambda\,{=}\,(\Omega_{\mathrm{t}}\cos\theta\,{+}\,\Omega_{\ell}\sin\theta)/\omega$. Equation (\ref{eq:H_tr1}) means that we can drive one-photon transitions at the degeneracy point via an antisymmetric magnetic field due to the~$\sin\theta$ term. This effect becomes weaker when we move away from the degeneracy point. However, in this case the longitudinal drive starts to activate transitions via the~$\cos\theta$-term. Additionally, there is a certain angle~$\theta^{\star}$, for which both drives cancel each other. In Figs.\,\ref{fig:Fig_S09}\,(a)\,--\,(d), we show numerical calculations of the one-photon transition using Eq.\,(\ref{eq:H_tr1}). Moving from (a) to (d), we increase the impact of the longitudinal drive, while staying in the low power (one-photon) limit, where~$\Omega_{\ell},\,\Omega_{\mathrm{t}}\,{\ll}\,\omega_{\mathrm{r}}$. As apparent, the coupling at the degeneracy point becomes weaker when increasing the ratio between longitudinal and transversal drive, implementing a controllable selection rule.\medskip

\textbf{Two-photon transitions} For increasing drive power, we can activate multi-photon transitions by shining electromagnetic fields of frequency~$\omega_{\mathrm{q}}/n$. In the case~$n\,{=}\,2$, the two photons have frequency $\omega\,{=}\,\omega_{\mathrm{q}}/2$ and a combined even parity. Therefore, two-photon processes are forbidden at the qubit degeneracy point for transversal and for longitudinal drives. For the general~$n$-photon case, we can express the parity as~$(\widehat{\Pi}_{\mathrm{q}}\hat{A}_{-}\widehat{\Pi}_{\mathrm{q}})^{n}\,{=}\,({-}1)^{n}\hat{A}_{-}^{n}$~\cite{Niemczyk_2011}. If we move away from the qubit degeneracy point, we can derive the effective Hamiltonian via a Schrieffer-Wolff transformation~\cite{Deppe_2008}. For possible two-photon transitions in the bare basis, i.e.,~$\omega\,{=}\,\Delta/\sin\theta$, we find
\begin{align}
 \widehat{\mathcal{H}}_{\mathrm{trans},2} &= \frac{\hbar}{2}\left[\frac{\Omega_{\ell}}{2}\cos\theta-\frac{\Omega_{\mathrm{t}}}{2}\sin\theta\right][-J_{1}(\lambda) - J_{3}(\lambda)]\hat{\sigma}_{x}\notag\\
 &\approx \frac{\hbar}{2}\left[\frac{\Omega_{\ell}}{2}\cos\theta-\frac{\Omega_{\mathrm{t}}}{2}\sin\theta\right]\left[-\frac{\Omega_{\mathrm{t}}}{2\omega}\cos\theta - \frac{\Omega_{\ell}}{2\omega}\sin\theta\right]\hat{\sigma}_{x}\notag\\
 &= \frac{\hbar}{8}\left[\left(\Omega_{\mathrm{t}}^{2} - \Omega_{\ell}^{2}\right)\sin^{2}\theta\cos\theta + \Omega_{\ell}\Omega_{\mathrm{t}}\left(\sin^{3}\theta - \cos^{2}\theta\sin\theta\right)\right]\hat{\sigma}_{x}\,.\label{eqn:2photon}
\end{align}
This Hamiltonian consists of two parts. First, a part proportional to~$\sin^{2}\theta\cos\theta$, which is well-known from circuit QED experiments studying the controlled symmetry breaking of two-photon processes~\cite{Deppe_2008,Niemczyk_2011}. For a mixed $(\Omega_{\ell}\,{+}\,\Omega_{\mathrm{t}})$ drive, there is an additional term proportional to $\Omega_{\ell}\Omega_{\mathrm{t}}$, which also depends strongly on the Bloch angle [see case $(\alpha)$ in Figs.\,\ref{fig:Fig_S09}\,(e)\,--\,(h)].\medskip

\textbf{Sidebands} In the one-photon case, the red sideband transition $(\omega_{\mathrm{q}}\,{-}\,\omega_{\mathrm{r}})$ and the blue sideband transition $(\omega_{\mathrm{q}}\,{+}\,\omega_{\mathrm{r}})$ are forbidden for a transversal $\hat{\sigma}_{x}$-like drive. However, due to the even parity of the~$\hat{\sigma}_{z}$ operator, sideband transitions are allowed for~$\hat{\sigma}_{z}$ drives~\cite{Blais_2007}. To calculate the transition matrix element, we chose a unitary transformation via
\begin{equation}
 \widehat{\mathcal{U}} = \exp\left[-\imath\frac{\Delta^{\prime}t}{2}\hat{\sigma}_{z}-\imath(\omega_{\mathrm{r}}t)\hat{a}^{\dagger}\hat{a}\right]\,,
\end{equation}
where~$\Delta^{\prime}\,{=}\,\Delta\,{+}\,(\gamma_{+}\,{+}\,\gamma_{-})/2$ with~$\gamma_{\pm}\,{=}\,g_{\mathrm{t}}/(\Delta\,{\pm}\,\omega_{\mathrm{r}})$. In this frame, within a RWA, the sideband Hamiltonians read~\cite{Billangeon_2015}
\begin{align}
 \widehat{\mathcal{H}}_{\mathrm{red}} &= \hbar g_{\mathrm{t}}(\gamma_{+}+\gamma_{-})\hat{\sigma}_{z}\hat{a}^{\dagger}\hat{a} - \frac{\hbar}{2}\left[\frac{\Omega_{\ell}}{2}\sin\theta-\frac{\Omega_{\mathrm{t}}}{2}\cos\theta\right]2\gamma_{-}[\hat{a}^{\dagger}\hat{\sigma}_{-} + \hat{a}\hat{\sigma}_{+}]\,,\\
\widehat{\mathcal{H}}_{\mathrm{blue}} &= \hbar g_{\mathrm{t}}(\gamma_{+}+\gamma_{-})\hat{\sigma}_{z}\hat{a}^{\dagger}\hat{a} - \frac{\hbar}{2}\left[\frac{\Omega_{\ell}}{2}\sin\theta-\frac{\Omega_{\mathrm{t}}}{2}\cos\theta\right]2\gamma_{+}[\hat{a}^{\dagger}\hat{\sigma}_{+} + \hat{a}\hat{\sigma}_{-}]\,.
\end{align}
As shown in Figs.\,\ref{fig:Fig_S09}\,(e)\,-\,(h), the transition probability for sideband transitions is opposite compared to the one-photon transition depicted in Figs.\,\ref{fig:Fig_S09}\,(a)\,-\,(f).

\end{document}